\documentclass[11pt,letterpaper]{article}

\pdfoutput=1

\PassOptionsToPackage{nosort}{cite}

\usepackage{amssymb}
\usepackage{amsfonts}
\usepackage{dsfont}
\usepackage{mathrsfs}
\usepackage{enumitem}
\usepackage[T1]{fontenc}
\usepackage[british,mcite]{chet}
\usepackage{tikz}
\usepackage{pgfplots}
\usepackage{float}
\usepackage{upgreek}
\usepackage{sansmath}
\usepackage[font=small,format=hang,labelfont={sf,bf}]{caption}

\mathcode`\0="7030 \mathcode`\1="7031 \mathcode`\2="7032 \mathcode`\3="7033
\mathcode`\4="7034 \mathcode`\5="7035 \mathcode`\6="7036 \mathcode`\7="7037
\mathcode`\8="7038 \mathcode`\9="7039

\usetikzlibrary{intersections,calc,arrows.meta,bending,patterns,decorations.pathmorphing,decorations.markings,colorbrewer}

\pgfplotsset{compat=1.18}

\definecolor{forestgreen}{rgb}{0.0, 0.27, 0.13}
\definecolor{darkblue}{rgb}{0.0, 0.0, 0.55}
\definecolor{darkred}{rgb}{0.55, 0.0, 0.0}
\hypersetup{colorlinks,
           linkcolor={forestgreen},
           citecolor={darkblue},
           urlcolor={darkred},
           pdftitle={Quantum Rotors on the Fuzzy Sphere and the Cubic CFT},
           pdfdisplaydoctitle
}

\newcommand{\lsp}{\hspace{0.5pt}}

\renewcommand{\geq}{\geqslant}
\renewcommand{\leq}{\leqslant}

\DeclareMathOperator{\diag}{diag}

\title{Quantum Rotors on the Fuzzy Sphere\\[5pt] and the Cubic CFT}

\author{Andreas Stergiou}

\affiliation{Department of Mathematics, King's College London, Strand, London WC2R 2LS, United Kingdom}

\date{April 2026}

\abstract{The three-dimensional cubic conformal field theory governs the critical behaviour of Heisenberg magnets with cubic anisotropy. Studying this theory non-perturbatively is challenging, because its most easily accessible observables are numerically very close to those of the more symmetric \(O(3)\) model. In this work, we overcome this difficulty using the fuzzy sphere regularisation method. By adding a cubic-invariant two-body interaction to the quantum rotor Hamiltonian used for the \(O(3)\) model, we break the continuous rotational symmetry by construction and unambiguously isolate the cubic critical point. Using exact diagonalisation and the density matrix renormalisation group, we calculate the scaling dimensions of several key operators, including the leading scalar singlets, and resolve the splitting of the \(O(3)\) rank-two traceless symmetric tensor into the \(E_g\) and \(T_{2g}\) representations of the cubic group. Our results are consistent with existing Monte Carlo, conformal perturbation theory, and \(\varepsilon\) expansion benchmarks, demonstrating the power of the fuzzy sphere in resolving closely spaced universality classes.}

\begin{document}

\maketitle

\section{Introduction}
Computational methods valid beyond perturbation theory are invaluable for understanding strongly interacting physical systems. In the context of interest for the present work, that of conformal field theory (CFT), the conformal bootstrap \cite{Rattazzi:2008pe, *Poland:2018epd} is a rigorous such framework that has had great success in determining physical quantities of CFTs, most notably the scaling dimensions of operators in the three-dimensional Ising model \cite{El-Showk:2012cjh, *Kos:2014bka, *Kos:2016ysd, *Chang:2024whx}. Recently, another promising and powerful approach for the study of CFTs at strong coupling has emerged, based on the idea of regularising the continuum theory by means of a discretised fuzzy sphere realisation. This approach, which beautifully leverages the state-operator correspondence, was initially applied to the three-dimensional Ising model in \cite{Zhu:2022gjc} and produced an impressive set of results matching those of the conformal bootstrap.

Both conformal bootstrap and fuzzy sphere methods have been applied to models beyond the 3D Ising model; see e.g.\ \cite{Zhou:2023qfi, *Zhou:2024zud, *Fan:2025bhc, *ArguelloCruz:2025zuq, *EliasMiro:2025msj, *He:2025ong, *Taylor:2025odf, *Zhou:2025rmv, *Voinea:2025iun, *Tang:2025wtj, *Huffman:2026qqq, *Dey:2026cso}. Of direct relevance to this work is the \(O(3)\) model, which has been studied using the conformal bootstrap in \cite{Kos:2013tga, *Kos:2015mba, Chester:2020iyt} and using the fuzzy sphere in \cite{PhysRevB.110.115113, *Guo:2025odn, Dey:2025zgn}. The agreement between the two approaches is again remarkable, with the fuzzy sphere method reproducing the bootstrap results to within a few percent. Given the manifest success of the fuzzy sphere method in capturing the essential physics of 3D CFTs, it is natural to ask whether it can be applied to other models of interest, particularly ones that are not as well understood as the Ising and \(O(N)\) models and have proven to be challenging to isolate for the conformal bootstrap. One such model is the cubic CFT, which is believed to describe the critical behaviour of Heisenberg magnets.

The physical relevance of the cubic CFT arises from the fact that cubic anisotropy is present in most magnetic materials. In the case of Heisenberg magnets, the associated lattice structure breaks the full \(O(3)\) rotational symmetry down to the discrete cubic group. The cubic group is the symmetry group of the cube, generated by reflections and \(90^\circ\) rotations about the Cartesian axes. It is given by \(\mathbb{Z}_2^{\,3}\rtimes S_3\), and it is also commonly denoted by \(O_h\), the group of full octahedral symmetry. The \(O(3)\to O_h\) breaking is captured in the Landau--Ginzburg framework by the addition of a cubic-symmetric perturbation to the isotropic \(O(3)\) Lagrangian \cite{Pelissetto:2000ek}. Whether the cubic perturbation is relevant or not, in the renormalisation group sense, determines whether the critical behaviour is described by the cubic fixed point or the \(O(3)\) fixed point. The latest numerical results from the conformal bootstrap \cite{Chester:2020iyt} indicate that the cubic fixed point is indeed the stable one, as was already suggested by the \(\varepsilon\) expansion \cite{Pelissetto:2000ek, Adzhemyan:2019gvv}. However, the difference in standard critical exponents between the cubic and \(O(3)\) universality classes is tiny, owing to the proximity of the two fixed points in renormalisation group space.

The conformal bootstrap is in principle capable of determining the critical exponents of any CFT, but it is not always easy to isolate the theory of interest. In the case of the cubic CFT, the cubic perturbation at the \(O(3)\) fixed point is very weakly relevant, and the cubic fixed point data are extremely close to the \(O(3)\) ones in the most easily accessible corners of the parameter space. Additionally, crossing equations derived with a particular symmetry in mind, e.g.\ cubic, might re-organise themselves to a symmetry-enhanced \(O(3)\) configuration, making it difficult to separate the cubic fixed point from the \(O(3)\) fixed point. These issues affected some early attempts to bootstrap the cubic theory \cite{Rong:2017cow,*Stergiou:2018gjj, *Kousvos:2018rhl, *Kousvos:2019hgc}. Excluding symmetry enhancement of the crossing equations has recently been addressed in \cite{Kousvos:2025ext} based on the notion of redundant operators, but access to sectors that enable the use of such operators required an enlargement of the four-point function system considered, which in turn made the problem significantly more computationally expensive.

This state of affairs makes the fuzzy sphere method rather attractive for the cubic CFT. Indeed, the Hamiltonian used in the fuzzy sphere method has the global symmetry of the sought theory hard-coded into it, and symmetry enhancement is not possible. Thus, one simply needs to consider a Hamiltonian with the proper symmetry and tune it to criticality to attain the cubic CFT. This is the aim of the present work, in which we rely heavily on the realisation of the \(O(3)\) model on the fuzzy sphere achieved in \cite{Dey:2025zgn}. That work used a truncated quantum rotor model \cite{Sachdev:2011fcc} to describe the \(O(3)\) model on the fuzzy sphere, and we will use a similar approach here. We will deform the Hamiltonian of \cite{Dey:2025zgn} by a suitable term, breaking the \(O(3)\) symmetry to cubic, and tune the Hamiltonian parameters to recover the cubic CFT. Similarly to \cite{Dey:2025zgn} and other fuzzy sphere studies, we will use both exact diagonalisation (ED) and density matrix renormalisation group (DMRG) \cite{White:1992zz, *White:1993zza} to compute the energy spectrum of our cubic fuzzy sphere Hamiltonian, the former for small system sizes and the latter for larger ones.

The most essential concern when applying the fuzzy sphere method is that of the effect of the finite system size on the obtained results. To address this issue, we will perform our computations for a range of parameters and system sizes, and attempt to extract results in the thermodynamic limit. The associated extrapolations are however non-rigorous, and we will be content with results lacking precise error estimates, but which we expect to be accurate to within a few percent. Whenever possible, we will compare with \(\varepsilon\) expansion, conformal perturbation theory, and Monte Carlo results found in the literature. While the fuzzy sphere finite-size effect is not carefully managed in this work, we still believe that the strongly-coupled information obtained represents accurately the true physical situation. Improvements in the handling of the finite-size effect would of course be highly desirable. The use of conformal generators \cite{Fardelli:2024qla, *Fan:2024vcz, *Fardelli:2026zas} could be helpful for this.

This paper is organised as follows. In the next section we describe in brief the truncated quantum rotor model realisation of the \(O(3)\) universality class achieved in \cite{Dey:2025zgn}, and introduce the cubic deformation that forms the basis of our results. In Section \ref{sec:results} we describe our method for obtaining the critical values of our Hamiltonian parameters, which follows the conformal perturbation theory approach of \cite{Dey:2025zgn}, and present our results with both ED and DMRG methods. All of our results have been obtained with the use of \href{https://www.fuzzified.world/}{\texttt{FuzzifiED}} \cite{Zhou:2025liv}. We conclude in Section \ref{sec:conclusion}.

\section{Fuzzy sphere Hamiltonians}\label{sec:Hamiltonian}
\subsection{Hamiltonian with \texorpdfstring{\(O(3)\)}{O(3)} symmetry}
The starting point for the construction of our Hamiltonian is the quantum rotor model \cite{Sachdev:2011fcc}. A rigid rotor is simply a three-dimensional rigid object. (We will exclusively consider the case of three dimensions.) The orientation of such an object in space requires three (Euler) angles to be specified. The special case of a linear rigid rotor, e.g.\ a dumbbell, requires only two angles to be specified. These angles define a point on an auxiliary two-sphere, \(S^2\), and every orientation of the linear rigid rotor can be mapped to a point on that \(S^2\). For \(N\) linear rigid rotors labelled by an index \(i\), \(i=1,\ldots,N\), each rotor can be assigned a three-vector \(\mathbf{n}_i\) with \(\mathbf{n}_i^2=1\). A quantum rotor is obtained when the three-vectors \(\mathbf{n}_i\) become operator-valued, \(\mathbf{n}_i\to \hat{\mathbf{n}}_i\), with the constraint \(\hat{\mathbf{n}}_i^2=\hat{\mathbb{I}}\), where \(\hat{\mathbb{I}}\) is the identity operator. If the momentum of the \(i\)-th rotor is denoted by \(\hat{\mathbf{p}}_i\), then the quantum rotor model is given by the Hamiltonian
\begin{equation}\label{eq:rotHam}
    \hat{H}_{\text{rotors}} = \frac{1}{2I}\sum_{i=1}^N \hspace{-2.5pt}\hat{\hspace{2.5pt}\mathbf{L}}_{i}^{\!\raisebox{-2pt}{\(\scriptstyle 2\)}} - J \sum_{\langle ij\rangle} \hat{\mathbf{n}}_i\cdot \hat{\mathbf{n}}_j\,,\qquad\hspace{-2.5pt}\hat{\hspace{2.5pt}\mathbf{L}}_{i}=\hat{\mathbf{n}}_i\times \hat{\mathbf{p}}_i\,,
\end{equation}
where \(I\) is the moment of inertia of the rotor, \(J\) a coupling constant, and \(\langle ij\rangle\) denotes a sum over nearest neighbours. The components of the angular momentum operator of the \(i\)-th rotor are given by \(\hat{L}_i^a=\epsilon^{abc}\hat{n}_i^b \hat{p}_i^c\) (no sum on \(i\)), with \(\epsilon^{abc}\) the fully antisymmetric symbol of three indices. Indices from the beginning of the Latin alphabet take the values \(x,y,z\) corresponding to the directions of three-dimensional space. The commutation relations among the components of \(\hspace{-2.5pt}\hat{\hspace{2.5pt}\mathbf{L}}_{i}\) and \(\hat{\mathbf{n}}_i\) are given by (no sum on \(i\))
\begin{subequations}
    \begin{align}
        [\hat{L}_i^a,\hat{L}_j^b]&=i\epsilon^{abc}\hat{L}_i^c\delta_{ij}\,,\\
        [\hat{L}_i^a,\hat{n}_j^b]&=i\epsilon^{abc}\hat{n}_i^c\delta_{ij}\,,\\
        [\hat{n}_i^a,\hat{n}_j^b]&=0\,,
    \end{align}
\end{subequations}
as follows from the standard commutation relations \([\hat{n}_i^a, \hat{p}_j^b]=i\delta_{ij}\delta^{ab}\hat{\mathbb{I}}\). The first term in \eqref{eq:rotHam} is the kinetic energy, while the second term is the interaction. By varying \(IJ\) in \eqref{eq:rotHam} we can realise a phase transition between a paramagnetic state, in which the kinetic term dominates, and a magnetically ordered state, in which the interaction term dominates \cite{Sachdev:2011fcc}.

The quantum rotor Hamiltonian \eqref{eq:rotHam} has \(O(3)\) symmetry. The operator \(\hspace{-2.5pt}\hat{\hspace{2.5pt}\mathbf{L}}_{i}^{\!\raisebox{-2pt}{\(\scriptstyle 2\)}}\) is the quadratic Casimir of the Lie algebra \(\mathfrak{so}(3)\) corresponding to the \(i\)-th rotor. With eigenstates of \(\hspace{-2.5pt}\hat{\hspace{2.5pt}\mathbf{L}}_{i}^{\!\raisebox{-2pt}{\(\scriptstyle 2\)}}\) and \(\hat{L}_i^z\) denoted by \(|l_i,m_i\rangle\), \(l_i=0,1,2,\ldots\,,\) \(m_i=-l_i,-l_i+1,\ldots,l_i-1,l_i\), we have
\begin{equation}
    \hspace{-2.5pt}\hat{\hspace{2.5pt}\mathbf{L}}_{i}^{\!\raisebox{-2pt}{\(\scriptstyle 2\)}}|l_i,m_i\rangle = l_i(l_i+1)|l_i,m_i\rangle\,,\qquad \hat{L}_i^z|l_i,m_i\rangle = m_i|l_i,m_i\rangle\,.
\end{equation} 
The kinetic energy in \eqref{eq:rotHam} can be written as
\begin{equation}
    \hat{H}_{\text{rotors}}^{\text{(kin)}} = \frac{1}{2I}\sum_{i=1}^N \hspace{-2.5pt}\hat{\hspace{2.5pt}\mathbf{L}}_{i}^{\!\raisebox{-2pt}{\(\scriptstyle 2\)}}=\frac{1}{2I}\sum_{i=1}^N\sum_{l_i=0}^{\infty}\sum_{m_i=-l_i}^{l_i}l_i(l_i+1)|l_i,m_i\rangle\langle l_i,m_i|\,.
\end{equation}
To obtain a finite-dimensional Hilbert space, we truncate the sum over \(l_i\) to a maximum value of \(1\) for each rotor. There are then four states for each rotor, organising themselves into a singlet and a triplet under the action of \(SO(3)\).

The states \(|0_i,0_i\rangle\), \(|1_i,-1_i\rangle\), \(|1_i,0_i\rangle\), \(|1_i,1_i\rangle\) for \(i=1,\ldots,N\) generate a basis of the Hilbert space of \(N\) rotors, and will be the basic ingredients in our construction of the second-quantised Hamiltonian that we will use in our numerical calculations. To take advantage of the regularisation prowess afforded to us by the fuzzy sphere, we will map the \(N\) rotors onto a system of \(N\) fermions moving on a sphere under the influence of the radial magnetic field created by a magnetic monopole at the centre of the sphere, and project to the lowest Landau level (LLL) \cite{Zhu:2022gjc}. Such a configuration was first studied in \cite{Haldane:1983xm}. To represent the truncated rotor model, which has four states per rotor, we follow \cite{Dey:2025zgn} and use a system where each fermion has four internal flavours. By applying a strong Hubbard-like repulsion and operating the system at quarter filling, we enforce a constraint where each orbital, which serves as the site on the fuzzy sphere, is occupied by exactly one fermion. The fermion flavours essentially carry the rotor's identity, with the first flavour corresponding to the rotor's \(l=0\) state, and the other three flavours representing the \(l=1\) states.

In the finite Hilbert space generated by the construction we just described, excited states are created by the action of fermionic creation operators \(\hat{c}_{l_i,m_i}^\dagger\) on the vacuum state \(|0\cdots 0\rangle\) (with \(4N\) zeroes). For example, to excite the third fermion to the \(|1_3,0_3\rangle\) state, we use \(\hat{c}_{1_3,0_3}^\dagger|0\cdots 0\rangle=|0\cdots 010\cdots0\rangle\), where there are 10 zeroes before the 1 and \(4N-11\) zeroes after it. If the filling fraction is not constrained, the dimension of the Hilbert space is \(2^{4N}\). At a filling fraction of \(1/4\), the dimension of the Hilbert space is \(2^{2N}\).

The Hamiltonian considered in \cite{Dey:2025zgn} has various contributions. There is a flavour-agnostic fermion repulsion term, \(\hat{H}_{\text{Hub}}\), which enforces single occupancy per LLL orbital. There is further a contribution, \(\hat{H}_{\text{Heis}}\), which captures the effects of the interaction term in \eqref{eq:rotHam}. This is translated into a four-fermion interaction between different-\(l\) orbitals, as follows from the fact that the orientation operator \(\hat{\mathbf{n}}_i\) connects the \(l=0\) and \(l=1\) states. This is the reason why both singlet and triplet flavours are required in the rotor truncation. Indeed, if we attempted to work only with the triplet, then \(\hat{H}_{\text{Heis}}\) would be trivial because the orientation operator would have no matrix elements at all within the triplet sector. Finally, there is a contribution, \(\hat{H}_{\text{trans}}\), that corresponds to the kinetic energy in \eqref{eq:rotHam} and assigns a lower energy to the first flavour (representing the \(SO(3)\) singlet) compared to the other three (representing the \(SO(3)\) triplet). By tuning the coefficient \(h\) of \(\hat{H}_{\text{trans}}\), a phase transition is obtained.

We will not describe in detail the derivation of the second-quantised, fuzzy sphere Hamiltonian of \cite{Dey:2025zgn} here. The reader is referred to \cite{Dey:2025zgn} for the details. The essential ingredient is a local density defined on the fuzzy sphere by
\begin{equation}
    n^I(\Omega)=\boldsymbol{\uppsi}^\dagger\mathcal{M}^I(\Omega)\boldsymbol{\uppsi}\,,
\end{equation}
expressed in terms of the fermions
\begin{equation}
    \boldsymbol{\uppsi}=\begin{pmatrix}
        \psi_{0,0}\\
        \psi_{1,-1}\\
        \psi_{1,0}\\
        \psi_{1,1}
    \end{pmatrix}
\end{equation}
and the matrices \(\mathcal{M}^I\). The matrices used in \cite{Dey:2025zgn} are \(\mathcal{M}^0=\diag(1,1,1,1)\), featuring in \(\hat{H}_{\text{Hub}}\), \(\mathcal{M}^{\mathcal{L}}=\diag(0,2,2,2)\), featuring in \(\hat{H}_{\text{trans}}\), and \(\mathcal{M}^{\mathcal{R}}\) featuring in \(\hat{H}_{\text{Heis}}\). The structure of \(\mathcal{M}^{\mathcal{R}}\) results in couplings between fermions from singlet and triplet flavours, as made explicit by the matrix elements given below.

After standard integrals involving monopole harmonics are performed and projection to the LLL takes place, we arrive at the second-quantised Hamiltonian
\begin{equation}\label{eq:O3Ham}
    \hat{H}_{O(3)}^{\text{LLL}}=\hat{H}_{\text{Hub}}^{\text{LLL}} + \hat{H}_{\text{Heis}}^{\text{LLL}} + \hat{H}_{\text{trans}}^{\text{LLL}}\,,
\end{equation}
where
\begin{subequations}
    \begin{align}
        \hat{H}_{\text{Hub}}^{\text{LLL}}&=u\sum_{i_1,i_2,i=-s}^s V_{i_1,i_2,i_2-i,i_1+i}\,\mathbf{c}_{i_1}^\dagger\mathbf{c}_{i_2}^\dagger\mathbf{c}_{i_1+i}\mathbf{c}_{i_2-i}\,,\\
        \hat{H}_{\text{Heis}}^{\text{LLL}}&=-v\sum_{i_1,i_2,i=-s}^s V_{i_1,i_2,i_2-i,i_1+i}\,\mathbf{c}_{i_1}^\dagger\mathbf{c}_{i_2}^\dagger\bar{\mathcal{M}}^{\mathcal{R}}\mathbf{c}_{i_1+i}\mathbf{c}_{i_2-i}\,,\\
        \hat{H}_{\text{trans}}^{\text{LLL}}&=h\sum_{i=-s}^s \mathbf{c}_i^\dagger \mathcal{M}^{\mathcal{L}}\mathbf{c}_i\,.
    \end{align}
\end{subequations}
Here \(s\) is the charge of the monopole that gives rise to the orbitals in which the fermionic operators
\begin{equation}
    \mathbf{c}_i^\dagger=\begin{pmatrix}
        c_{0_i,0_i}^\dagger & c_{1_i,-1_i}^\dagger & c_{1_i,0_i}^\dagger & c_{1_i,1_i}^\dagger
    \end{pmatrix},\qquad
    \mathbf{c}_i=\begin{pmatrix}
        c_{0_i,0_i}\\
        c_{1_i,-1_i}\\
        c_{1_i,0_i}\\
        c_{1_i,1_i}
    \end{pmatrix}
\end{equation}
can create and annihilate excitations, respectively. The number of rotors \(N\) is determined by \(s\) via the relation \(N=2s+1\). We will typically take \(s\) to be a half-integer, and our rotors will be represented by the fermionic operators \(\mathbf{c}_i\) with half-integer labels from now on. The integrals over the fuzzy sphere determine \cite{Zhu:2022gjc}
\begin{equation}
    V_{i_1,i_2,i_3,i_4}=\sum_{\sigma=0}^{2s} V_\sigma\lsp(4s-2\sigma+1)\begin{pmatrix}
        s & s & 2s-\sigma\\
        i_1 & i_2 & -i_1-i_2
    \end{pmatrix} \begin{pmatrix}
        s & s & 2s-\sigma\\
        i_3 & i_4 & -i_3-i_4
    \end{pmatrix},
\end{equation}
where \(\begin{pmatrix} s_1 & s_2 & s_3 \\ i_1 & i_2 & i_3\end{pmatrix}\) is a Wigner \(3j\) coefficient. Finally, the matrix elements \(\bar{\mathcal{M}}^{\mathcal{R}}\), which essentially define \(\mathcal{M}^{\mathcal{R}}\), are given by
\begin{equation}
    \bar{\mathcal{M}}^{\mathcal{R}}=\sum_{\alpha=-1,0,1}(-1)^\alpha\lsp\langle l_1,m_1;l_2,m_2|\hat{n}_i^\alpha\hat{n}_j^{-\alpha}|l_3,m_3;l_4,m_4\rangle\,,
\end{equation}
where the spherical components \(\hat{n}^\alpha\) of the orientation operator are defined in terms of its Cartesian components by
\begin{equation}
    \hat{n}_i^0=\hat{n}_i^z\,,\qquad \hat{n}_i^{\pm 1}=\mp\frac{1}{\sqrt{2}}(\hat{n}_i^x\pm i\hat{n}_i^y)\,.
\end{equation}
Throughout this work, indices \(\alpha,\beta,\ldots\) from the beginning of the Greek alphabet label spherical components and take the values \(-1,0,1\), while indices \(a,b,\ldots\) from the beginning of the Latin alphabet label Cartesian components and take the values \(x,y,z\). The two sets of components are related by \((\hat{n}^\alpha)^\dagger=(-1)^\alpha\lsp\hat{n}^{-\alpha}\), so that the \(O(3)\)-invariant scalar product of the orientation operators of two rotors reads
\begin{equation}\label{eq:sphcontraction}
    \sum_{\alpha=-1,0,1}(-1)^\alpha\lsp \hat{n}_i^\alpha\lsp \hat{n}_j^{-\alpha} = \sum_{a=x,y,z}\hat{n}_i^a\lsp\hat{n}_j^a\,.
\end{equation}
It is the contraction \eqref{eq:sphcontraction}, with the index \(\alpha\) appearing with opposite signs on the two operators, that is required here; the naive expression \(\sum_\alpha \hat{n}_i^\alpha \hat{n}_j^\alpha=\hat{n}_i^x\hat{n}_j^x-\hat{n}_i^y\hat{n}_j^y+\hat{n}_i^z\hat{n}_j^z\) is not \(O(3)\) invariant. The matrix elements evaluate to \cite{Dey:2025zgn}
\begin{equation}
    \begin{aligned}
        \bar{\mathcal{M}}^{\mathcal{R}}_{l_1,m_1;l_2,m_2|l_3, m_3;l_4,m_4} &= \sum_{\alpha=-1,0,1}(-1)^{-m_1-m_2-\alpha}\sqrt{(2l_1+1)(2l_2+1)(2l_3+1)(2l_4+1)}\\
        &\qquad\qquad \begin{pmatrix} l_1 & 1 & l_4\\ 0 & 0 & 0\end{pmatrix} \begin{pmatrix} l_2 & 1 & l_3\\ 0 & 0 & 0\end{pmatrix} \begin{pmatrix} l_1 & 1 & l_4\\ -m_1 & -\alpha & m_4\end{pmatrix} \begin{pmatrix} l_2 & 1 & l_3\\ -m_2 & \alpha & m_3\end{pmatrix}.
    \end{aligned}   
\end{equation}
These ingredients allow us to fully specify the Hamiltonian that follows from \(\hat{H}_{\text{O(3)}}^{\text{LLL}}\) of \eqref{eq:O3Ham} as a \(2^{2N}\times 2^{2N}\) matrix of numbers, after we assign values to \(u,v,h\) and \(V_0,\ldots,V_{2s}\).

\subsection{Hamiltonian with cubic symmetry}
To obtain a Hamiltonian with cubic symmetry, we must add to the \(O(3)\)-symmetric Hamiltonian \eqref{eq:O3Ham} a term that breaks the continuous \(O(3)\) symmetry down to the discrete cubic group \(O_h\). A natural cubic-invariant deformation couples the squared Cartesian components of the orientation operators. Schematically, this interaction takes the form
\begin{equation}\label{eq:cubicdef}
    \sum_{a=x,y,z} (\hat{n}^a)^2 \otimes (\hat{n}^a)^2\,,
\end{equation}
where the tensor product indicates a two-body coupling between rotors and, as above, \(a\) runs over the Cartesian values \(x,y,z\). In this notation, \(\hat{H}_{\text{Heis}}\) is schematically of the form \(\sum_a \hat{n}^a\otimes\hat{n}^a\), which is the Cartesian form of the invariant spherical contraction \eqref{eq:sphcontraction}. The interaction \eqref{eq:cubicdef} manifestly respects cubic symmetry while breaking the full \(O(3)\) invariance. Note that the \(O(3)\)-symmetric combination \(\sum_{a}(\hat{n}^a)^2 = \hat{\mathbf{n}}^2 = \hat{\mathbb{I}}\) is trivial, but the individual \((\hat{n}^a)^2\) terms are not.

To build a geometric intuition for this deformation, consider a semi-classical limit where adjacent rotors align, \(\hat{\mathbf{n}}_i \approx \hat{\mathbf{n}}_j \approx \mathbf{n}\). The two-body interaction then reduces to an effective single-rotor potential proportional to \(n_x^4 + n_y^4 + n_z^4\). For a unit vector \(\mathbf{n}\) confined to the sphere, this potential is highly anisotropic. It exhibits maxima along the Cartesian axes, corresponding to the six faces of a cube, and minima along the body diagonals, corresponding to the eight corners. By reshaping the previously featureless \(O(3)\) energy landscape into this ``bumpy'' surface, the deformation explicitly breaks the continuous rotational symmetry and energetically pins the rotors to the discrete orientations of a cubic lattice.

On the fuzzy sphere, this deformation is realised by adding to the LLL Hamiltonian a term with the same structure as \(\hat{H}_{\text{Heis}}^{\text{LLL}}\), but with the flavour matrix \(\bar{\mathcal{M}}^{\mathcal{R}}\) replaced by a matrix \(\bar{\mathcal{M}}^{\mathcal{C}}\) encoding the cubic anisotropy. To construct \(\bar{\mathcal{M}}^{\mathcal{C}}\), we must determine the matrix elements of \((\hat{n}^a)^2\) in the truncated rotor basis. Recall that the orientation operator \(\hat{n}^a\) connects singlet and triplet states, with matrix elements given by
\begin{equation}
    N^a_{l_1,m_1|l_2,m_2} = \langle l_1,m_1|\hat{n}^a|l_2,m_2\rangle\,,
\end{equation}
which are non-vanishing only when \(|l_1-l_2|=1\). In the truncated basis \(\{|0,0\rangle, |1,-1\rangle, |1,0\rangle, |1,1\rangle\}\), the matrices \(N^a\) are \(4\times 4\) and take the form
\begin{equation}
    N^x = \frac{1}{\sqrt{6}}\begin{pmatrix}
        0 & 1 & 0 & -1\\
        1 & 0 & 0 & 0\\
        0 & 0 & 0 & 0\\
        -1 & 0 & 0 & 0
    \end{pmatrix}, \qquad
    N^y = \frac{i}{\sqrt{6}}\begin{pmatrix}
        0 & -1 & 0 & -1\\
        1 & 0 & 0 & 0\\
        0 & 0 & 0 & 0\\
        1 & 0 & 0 & 0
    \end{pmatrix},\qquad
    N^z = \frac{1}{\sqrt{3}}\begin{pmatrix}
        0 & 0 & 1 & 0\\
        0 & 0 & 0 & 0\\
        1 & 0 & 0 & 0\\
        0 & 0 & 0 & 0
    \end{pmatrix}.
\end{equation}
The orientation operator enters the Hamiltonian only through the combination \(v\,\hat{n}^a\otimes\hat{n}^a\), and may therefore be rescaled at will provided the coupling \(v\) is rescaled to compensate. The implementation accompanying this work follows \cite{Dey:2025zgn} in using a normalisation of \(\hat{n}^a\) that differs from the matrix elements above by an overall constant, which is absorbed into the quoted value of \(v\). 

It should be stressed that the matrix elements of \((\hat{n}^a)^2\) in the truncated basis are \emph{not} obtained by squaring the matrices \(N^a\). Denoting by \(P_{\rm tr}\) the projector onto the truncated space, the square \((N^a)^2\) computes \(P_{\rm tr}\hat{n}^aP_{\rm tr}\hat{n}^aP_{\rm tr}\), whereas the object required in \eqref{eq:cubicdef} is \(P_{\rm tr}(\hat{n}^a)^2P_{\rm tr}\). The two are not equal, because \(\hat{n}^a\) acting on an \(l=1\) state produces components with \(l=0\) and \(l=2\), and the intermediate projection discards the latter. (In the singlet sector the two do agree, since \(\hat{n}^a|0,0\rangle\) lies entirely within \(l=1\).) We therefore compute \(P_{\rm tr}(\hat{n}^a)^2P_{\rm tr}\) directly below.

It is convenient to first introduce projectors onto the Cartesian directions within the triplet sector. Defining the Cartesian triplet states
\begin{equation}
    |x\rangle = \frac{1}{\sqrt{2}}\bigl(|1,-1\rangle-|1,1\rangle\bigr)\,,\qquad |y\rangle = \frac{i}{\sqrt{2}}\bigl(|1,-1\rangle+|1,1\rangle\bigr)\,,\qquad |z\rangle = |1,0\rangle\,,
\end{equation}
the projectors \(P^a = |a\rangle\langle a|\) satisfy \(\sum_a P^a=\Pi_1\), with \(\Pi_1\) the projector onto the triplet sector, and provide a natural representation of the cubic anisotropy restricted to the triplet. Schematically, this takes the form
\begin{equation}\label{eq:projcubic}
    \sum_{a=x,y,z} P^a \otimes P^a\,,
\end{equation}
where the explicit projector matrices in the basis \(\{|1,-1\rangle, |1,0\rangle, |1,1\rangle\}\) are
\begin{equation}
    P^x = \frac12\begin{pmatrix}
        1 & 0 & -1\\
        0 & 0 & 0\\
        -1 & 0 & 1
    \end{pmatrix}, \qquad
    P^y = \frac12\begin{pmatrix}
        1 & 0 & 1\\
        0 & 0 & 0\\
        1 & 0 & 1
    \end{pmatrix}, \qquad
    P^z = \begin{pmatrix}
        0 & 0 & 0\\
        0 & 1 & 0\\
        0 & 0 & 0
    \end{pmatrix}.
\end{equation}
The formulation \eqref{eq:projcubic} acts purely within the triplet sector and does not introduce any additional singlet-triplet mixing beyond what is already present in \(\hat{H}_{\text{Heis}}^{\text{LLL}}\).

We may now write down the correctly truncated matrix elements of \((\hat{n}^a)^2\). The singlet-triplet block vanishes by parity, \(\langle 0,0|(\hat{n}^a)^2|1,m\rangle\propto\int_{S^2}n^an^an^b\,d\Omega=0\), so that \(P_{\rm tr}(\hat{n}^a)^2P_{\rm tr}\) is block diagonal. The singlet element is fixed by rotational invariance, \(\langle 0,0|(\hat{n}^a)^2|0,0\rangle=\frac13\). Within the triplet, the Cartesian states have wavefunctions \(\psi_b(\mathbf{n})=\sqrt{3/4\pi}\lsp n^b\), and using \(\int_{S^2}n^bn^an^an^c\,d\Omega=\frac{4\pi}{15}(\delta^{bc}+2\delta^{ab}\delta^{ac})\), with no sum on \(a\), one finds \(\langle b|(\hat{n}^a)^2|c\rangle=\frac15\delta^{bc}+\frac25\delta^{ab}\delta^{ac}\). Hence, with \(\Pi_0\) and \(\Pi_1\) the projectors onto the singlet and triplet sectors,
\begin{equation}\label{eq:Qa}
    Q^a=P_{\rm tr}\lsp(\hat{n}^a)^2\lsp P_{\rm tr} = \tfrac13\lsp\Pi_0+\tfrac15\lsp\Pi_1+\tfrac25\lsp P^a\,.
\end{equation}
The constraint \(\sum_a(\hat{n}^a)^2=\hat{\mathbb{I}}\) is satisfied by \eqref{eq:Qa} exactly, \(\sum_a Q^a=\mathds{1}\), in both the singlet and the triplet sectors, and no regularisation of the truncated operator is required. Here and below \(\mathds{1}\) denotes the identity matrix on the four-dimensional truncated space, as opposed to the identity operator \(\hat{\mathbb{I}}\) on the full rotor Hilbert space. By contrast, squaring the matrices \(N^a\) gives \(\frac13\Pi_0+\frac13 P^a\), whose sum over \(a\) equals \(\mathds{1}\) in the singlet sector but only \(\frac13\mathds{1}\) in the triplet sector; the deficit is precisely the weight carried by the \(l=2\) intermediate states removed by the intermediate projection.

The cubic interaction built from \eqref{eq:Qa} is now readily decomposed. Writing \(Q^a=A+\frac25 P^a\), where \(A=\frac13\Pi_0+\frac15\Pi_1=\frac13\mathds{1}-\frac{2}{15}\Pi_1\) is an \(O(3)\) singlet, and using \(\sum_a P^a=\Pi_1\), we obtain
\begin{equation}\label{eq:Qdecomp}
    \sum_{a=x,y,z}Q^a\otimes Q^a = \frac{4}{25}\sum_{a=x,y,z}P^a\otimes P^a + 3\lsp A\otimes A + \frac25\lsp\bigl(A\otimes\Pi_1+\Pi_1\otimes A\bigr).
\end{equation}
Every term on the right-hand side of \eqref{eq:Qdecomp} other than the first is \(O(3)\) invariant: both \(\mathds{1}=\mathcal{M}^0\) and \(\Pi_1=\frac12\mathcal{M}^{\mathcal{L}}\) are \(O(3)\)-invariant flavour matrices, so these terms are ordinary density-density two-body interactions and cannot generate any cubic anisotropy. Their only effect is to displace the Hamiltonian along \(O(3)\)-invariant directions; of these, the single relevant singlet direction is retuned by \(h\) in Section \ref{sec:results}, while the remaining ones are irrelevant and do not affect the universality class. The entire \(O_h\)-breaking content of \(\sum_a Q^a\otimes Q^a\) is therefore carried by \(\sum_a P^a\otimes P^a\), and the two constructions are equivalent up to the rescaling \(w\to\frac{4}{25}w\) of the cubic coupling together with a shift of the \(O(3)\)-symmetric part of the Hamiltonian. We accordingly adopt \eqref{eq:projcubic} as the definition of the cubic deformation, with \(w\) the associated coupling, in all that follows.

The full cubic Hamiltonian on the fuzzy sphere is then
\begin{equation}\label{eq:cubicHam}
    \hat{H}_{\text{cubic}}^{\text{LLL}} = \hat{H}_{O(3)}^{\text{LLL}} + w\sum_{i_1,i_2,i=-s}^s V_{i_1,i_2,i_2-i,i_1+i}\,\mathbf{c}_{i_1}^\dagger\mathbf{c}_{i_2}^\dagger\bar{\mathcal{M}}^{\mathcal{C}}\mathbf{c}_{i_1+i}\mathbf{c}_{i_2-i}\,,
\end{equation}
where \(\bar{\mathcal{M}}^{\mathcal{C}}\) encodes the cubic anisotropy. When using the projector formulation \eqref{eq:projcubic}, we have
\begin{equation}
    \bar{\mathcal{M}}^{\mathcal{C}}_{l_1,m_1;l_2,m_2|l_3,m_3;l_4,m_4} = \sum_{a=x,y,z} (P^a)_{m_1,m_4}(P^a)_{m_2,m_3}\,,
\end{equation}
with the understanding that the projectors act only within the triplet sector and vanish when any of the indices corresponds to the singlet state.

\section{Spectrum results at criticality}\label{sec:results}
\subsection{Summary of pre-existing results}
Precise values of operator dimensions for the cubic theory in \(d=3\) are not available for a wide class of operators. Various results have been reported in the literature over the years. The most comprehensive treatment of the operator spectrum of the cubic theory was performed within the \(\varepsilon\) expansion in \cite{Antipin:2019vdg, *Bednyakov:2023lfj, *Henriksson:2025hwi, *Henriksson:2025vyi}. 

The state-of-the-art non-perturbative Monte Carlo study of \cite{Hasenbusch:2022zur} has determined 
\begin{equation}
    \Delta_\phi=0.51891(7)\,,\qquad \Delta_S=1.5937(6)\,,\qquad \Delta_{S'}=3.0133(8)\,.
\end{equation}
A fixed-dimension expansion calculation at six loops \cite{Carmona:1999rm} has given 
\begin{equation}
    \Delta_\phi=0.5166(13)\,,\qquad \Delta_S=1.584(12)\,,\qquad \Delta_{S'}=3.010(4)\,,
\end{equation}
while the \(\varepsilon\) expansion at six loops \cite{Adzhemyan:2019gvv} has suggested, after a Pad\'e--Borel--Leroy resummation,
\begin{equation}
    \Delta_\phi=0.5180(15)\,,\qquad \Delta_S=1.571(16)\,,\qquad \Delta_{S'}=3.005(5)\,.
\end{equation}
Conformal perturbation theory \cite{Rong:2023owx} has given
\begin{equation}
    \Delta_X=1.2256(36)\,,\qquad \Delta_Z=1.1988(24)\,,\qquad \Delta_{S'}=3.0115(21)\,,
\end{equation}
which includes the non-singlet operators $X$ and $Z$ that emerge from the decomposition of the leading rank-two traceless symmetric operator of $O(3)$ under the cubic subgroup.

\subsection{Tuning to criticality}
All calculations presented in this work were performed with the use of \href{https://www.fuzzified.world/}{\texttt{FuzzifiED}} \cite{Zhou:2025liv}, and rely on an extension of the example file \texttt{o3\_wf\_spectrum.jl} found at \url{https://docs.fuzzified.world/tutorial/#List-of-Examples}. The \href{https://julialang.org/}{\texttt{Julia}} code used in this work can be found in a dedicated \href{https://github.com/}{\texttt{GitHub}} repository, \href{https://github.com/andstergiou/fuzzy-cubic}{\texttt{andstergiou/fuzzy-cubic}}, which is archived on \href{https://zenodo.org/}{\texttt{Zenodo}}~\cite{fuzzy_cubic_code}.

ED at \(N=12\) is feasible, but each of the 16 sectors requires approximately 600 GB of memory, working with a Hilbert space of dimension \(2^{20}=1,048,576\). Going up by a factor of 16 to the Hilbert space dimension of the \(N=14\) case is not feasible, but DMRG can comfortably obtain low-lying states for that \(N\) and even higher. We have used DMRG up to \(N=22\).

Our first task is to choose the values of the various parameters in our Hamiltonian. Here we will follow \cite{Dey:2025zgn} and fix
\begin{equation}
    V_0=6.5\,,\qquad V_1=1.0\,,\qquad V_{\sigma>1}=0\,,\qquad u=1.0\,,\qquad v=1.4\,.
\end{equation}
We further need to fix the values of \(w\) and \(h\). We have explored the range \(w=0.2,0.4,0.6,0.8\) and, after comparing the quality of the spectra across these values, we have identified 
\begin{equation}
    w=0.6
\end{equation}
as our preferred choice; it is this value on which we focus in the remainder of the paper. For each value of \(w\), the corresponding \(h\) at a given \(N\) is determined by the conformal perturbation theory strategy of \cite{Dey:2025zgn}, which we now describe.

When the Hamiltonian is detuned from the critical value $h_c$ by a small amount \(\delta h = h-h_c\), the leading perturbation to the spectrum is generated by the relevant scalar \(S\) of the critical theory, with effective coupling \(g_S(h)\). The energy gap of a state associated with a primary operator \(\mathcal{O}\), measured from the vacuum, takes the form \cite{Dey:2025zgn}
\begin{equation}\label{eq:cpt}
    \delta E_{\mathcal{O}} = \frac{c}{R}\,\Delta_{\mathcal{O}} + g_S\, f_{\mathcal{O}S\mathcal{O}}\,,
\end{equation}
where \(c\) is the speed of light of the fuzzy sphere regularisation, \(R\propto\sqrt{N}\) is the radius of the sphere, \(\Delta_{\mathcal{O}}\) is the scaling dimension of \(\mathcal{O}\), and \(f_{\mathcal{O}S\mathcal{O}}\) is an appropriate combination of OPE coefficients and kinematic factors. Using two distinct states yields a \(2\times 2\) linear system for \(c/R\) and \(g_S\). Following \cite{Dey:2025zgn}, we use the fundamental scalar \(\phi\) and its first descendant \(\partial_\mu\phi\), with scaling dimensions \(\Delta_\phi\approx 0.5189\) and \(\Delta_{\partial_\mu\phi}=1+\Delta_\phi\approx 1.5189\). On the fuzzy sphere these states are identified by their total angular momentum: \(\phi\) has \(L=0\) and \(\partial_\mu\phi\) has \(L=1\). In the $O(3)$ case, the OPE data entering~\eqref{eq:cpt} for these two states take the values \(f_{\phi S\phi}\approx 0.525\) and \(f_{\partial_\mu\phi\,S\,\partial_\mu\phi}\approx 0.083\), as determined from the \(O(3)\) conformal bootstrap \cite{Dey:2025zgn}. The linear system is
\begin{equation}\label{eq:linsys}
    \begin{pmatrix} \Delta_\phi & f_{\phi S\phi} \\ \Delta_{\partial_\mu\phi} & f_{\partial_\mu\phi\,S\,\partial_\mu\phi} \end{pmatrix} \begin{pmatrix} c/R \\ g_S \end{pmatrix} = \begin{pmatrix} \delta E_\phi \\ \delta E_{\partial_\mu\phi} \end{pmatrix},
\end{equation}
from which \(g_S(h)\) is extracted for each value of \(h\). The optimal value \(h_{\rm opt}\) is the root \(g_S(h_{\rm opt})=0\), found by the secant method.

Since \eqref{eq:cpt} must of course be applied with the data of the cubic rather than those of the \(O(3)\) theory, it is worth spelling out which CFT data actually control \(h_{\rm opt}\). Solving \eqref{eq:linsys} gives
\begin{equation}\label{eq:gSsol}
    g_S = \frac{\Delta_\phi\,\delta E_{\partial_\mu\phi}-\Delta_{\partial_\mu\phi}\,\delta E_\phi}{\Delta_\phi\, f_{\partial_\mu\phi\,S\,\partial_\mu\phi}-\Delta_{\partial_\mu\phi}\,f_{\phi S\phi}}\,,
\end{equation}
so that the OPE data enter only through the overall normalisation of \(g_S\) and drop out of the condition \(g_S(h_{\rm opt})=0\) entirely. The tuning criterion is simply that the two gaps be in the ratio predicted by the CFT,
\begin{equation}\label{eq:tunecrit}
    \frac{\delta E_{\partial_\mu\phi}}{\delta E_\phi}\bigg|_{h=h_{\rm opt}} = \frac{\Delta_{\partial_\mu\phi}}{\Delta_\phi} = 1+\frac{1}{\Delta_\phi}\,,
\end{equation}
and the values quoted above for \(f_{\phi S\phi}\) and \(f_{\partial_\mu\phi\,S\,\partial_\mu\phi}\), which are available only for the \(O(3)\) theory, have no bearing on the extracted \(h_{\rm opt}\). The only CFT input is thus \(\Delta_\phi\), for which we use the cubic Monte Carlo value \(\Delta_\phi=0.51891(7)\) of \cite{Hasenbusch:2022zur}; the corresponding \(O(3)\) value differs from it by roughly one part in \(10^4\) \cite{Chester:2020iyt}, so that the residual systematic incurred by the proximity of the two fixed points is entirely negligible at the present level of accuracy.

\begin{figure}[H]
    \centering
    \begin{tikzpicture}
        \begin{axis}[
            font=\sffamily\sansmath,
            width=0.55\textwidth,
            height=0.42\textwidth,
            xlabel={\(N\)},
            ylabel={\(h_{\rm opt}(N)\)},
            xmin=9.5, xmax=26.5,
            ymin=14.88, ymax=15.42,
            xtick={10,12,14,16,18,20,22,24,26},
            ytick distance=0.1,
            y tick label style={/pgf/number format/.cd, fixed, fixed zerofill, precision=1},
            grid=both,
            grid style={line width=0.2pt, draw=gray!30},
            major grid style={line width=0.4pt, draw=gray!40},
            tick align=inside,
            legend style={at={(0.97,0.96)},anchor=north east,font=\sffamily\sansmath\small},
            legend cell align=left,
            ]
            \addplot[thick, orange!80!black, smooth, forget plot] coordinates {
                (10.0000,15.45560) (10.4051,15.42307) (10.8101,15.39378)
                (11.2152,15.36730) (11.6203,15.34328) (12.0253,15.32141)
                (12.4304,15.30144) (12.8354,15.28315) (13.2405,15.26635)
                (13.6456,15.25088) (14.0506,15.23660) (14.4557,15.22338)
                (14.8608,15.21113) (15.2658,15.19974) (15.6709,15.18914)
                (16.0759,15.17925) (16.4810,15.17001) (16.8861,15.16137)
                (17.2911,15.15326) (17.6962,15.14565) (18.1013,15.13850)
                (18.5063,15.13176) (18.9114,15.12541) (19.3165,15.11941)
                (19.7215,15.11375) (20.1266,15.10839) (20.5316,15.10331)
                (20.9367,15.09850) (21.3418,15.09393) (21.7468,15.08958)
                (22.1519,15.08545) (22.5570,15.08152) (22.9620,15.07778)
                (23.3671,15.07421) (23.7722,15.07080) (24.1772,15.06754)
                (24.5823,15.06443) (24.9873,15.06145) (25.3924,15.05860)
                (25.7975,15.05587) (26.0000,15.05455)
            };
            \addplot[thick, orange!80!black] coordinates {(0,0)(0,0)};
            \addlegendentry{\(h_c+A/N^{b/2}\)}
            \addplot[dashed, thick, green!40!black, domain=9.5:26.5, samples=2, forget plot] {14.955354};
            \node[green!40!black, anchor=south west] at (axis cs:9.6,14.955354) {\(h_c\)};
            \addplot[only marks, mark=*, mark size=2pt, green!40!black]
              coordinates {(12,15.32264) (14,15.23856) (16,15.18070) (18,15.14038)};
            \addlegendentry{DMRG data}
            \addplot[only marks, mark=*, mark size=2pt, draw=green!40!black, fill=white]
              coordinates {(20,15.11003) (22,15.08698)};
            \addlegendentry{Extrapolated}
        \end{axis}
    \end{tikzpicture}
    \caption{Finite-size dependence of \(h_{\rm opt}(N)\) for \(w=0.6\). Filled circles are the DMRG values obtained via the secant method applied to \(g_S(h_{\rm opt})=0\); open circles are extrapolated using the fit~\eqref{eq:hfit} with \eqref{eq:hfitval}. The orange curve is the three-parameter fit with \eqref{eq:hfitval}; the dashed line marks \(h_c=14.955(9)\).}
\label{fig:hopt}
\end{figure}
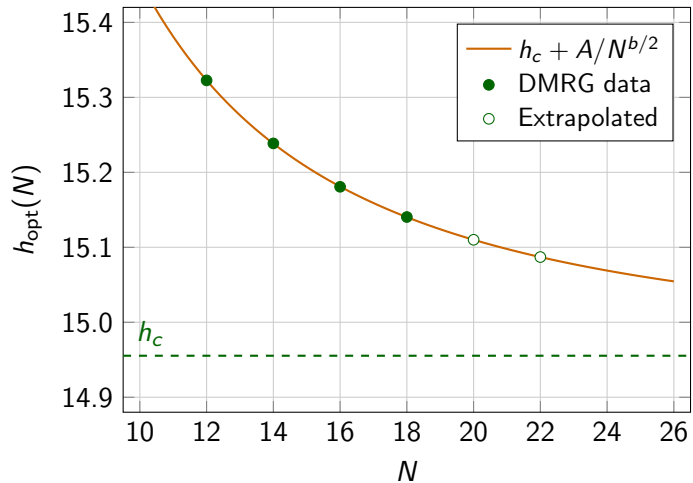

Since the critical point acquires finite-size corrections, \(h_{\rm opt}\) depends on \(N\). This dependence is well described by
\begin{equation}\label{eq:hfit}
    h_{\rm opt}(N) = h_c + \frac{A}{N^{b/2}}\,,
\end{equation}
with parameters \(h_c\), \(A\), and \(b\) fit to the numerical data. Focusing on our preferred choice \(w=0.6\), a least-squares fit to the DMRG data at \(N=12,14,16,18\) yields the parameters
\begin{equation}\label{eq:hfitval}
    h_c = 14.955(9)\,,\qquad A = 24.7\pm 3.0\,,\qquad b = 3.39(12)\,,
\end{equation}
where the uncertainties correspond to the standard errors of the fit parameters. The resulting values of \(h_{\rm opt}(N)\) used in all subsequent calculations are collected in Table~\ref{tab:hopt}, and the corresponding finite-size dependence is illustrated in Fig.~\ref{fig:hopt}. The figure confirms that the finite-size ansatz~\eqref{eq:hfit} describes the data well and that \(h_{\rm opt}(N)\) converges smoothly towards \(h_c\) as \(N\) grows. 

\begin{table}[H]
\centering
\begin{tabular}{c|cccccc}
    \(N\) & 12 & 14 & 16 & 18 & 20 & 22 \\\hline
    \(h_{\rm opt}(N)\) & 15.323 & 15.239 & 15.181 & 15.140 & 15.110 & 15.087 \\
\end{tabular}
\caption{Optimal values \(h_{\rm opt}(N)\) of the transverse-field parameter for each system size \(N\) at \(w=0.6\). Values for \(N\leq 18\) are obtained directly from the DMRG optimisation; values for \(N=20,22\) are extrapolated using the fit~\eqref{eq:hfit}.}
\label{tab:hopt}
\end{table}

The value \(h_c=14.955(9)\) obtained at \(w=0.6\) lies close to the value \(h_c^{O(3)}=14.992\) reported in \cite{Dey:2025zgn} for the \(O(3)\) Hamiltonian, i.e.\ for \(w=0\). This proximity is expected rather than suspicious, and it can be checked directly by repeating the analysis at each of the values of \(w\) we have explored. Applying the fit \eqref{eq:hfit} to the \(N=12,14,16,18\) data at each \(w\) gives the critical values collected in Table~\ref{tab:hcw}.

\begin{table}[H]
\centering
\begin{tabular}{c|cccc}
    \(w\) & 0.2 & 0.4 & 0.6 & 0.8 \\\hline
    \(h_c(w)\) & 14.985 & 14.957 & 14.955 & 14.855 \\
\end{tabular}
\caption{Critical values \(h_c(w)\) obtained from the fit \eqref{eq:hfit} to the DMRG data at \(N=12,14,16,18\) for each value of the cubic coupling \(w\). The uncertainties are comparable to the \(\pm0.009\) quoted in \eqref{eq:hfitval} for \(w=0.6\).}
\label{tab:hcw}
\end{table}
\noindent
Two features of Table~\ref{tab:hcw} are relevant here. First, \(h_c(w)\) decreases monotonically with increasing \(w\), so that the displacement of the critical point away from the \(O(3)\) value is a systematic effect of the cubic deformation rather than numerical scatter. Second, and more directly, at the smallest coupling explored we obtain \(h_c(0.2)=14.985\), which agrees with the \(O(3)\) determination \(h_c^{O(3)}=14.992\) of \cite{Dey:2025zgn} to better than \(0.05\%\) and well within the fit uncertainty. Since the Hamiltonian \eqref{eq:cubicHam} reduces to the \(O(3)\) Hamiltonian as \(w\to0\), this is exactly the behaviour required for the determination of \(h_c\) to be trustworthy, and it provides an independent check of the whole tuning procedure. We do not attempt a numerical extrapolation of \(h_c(w)\) to \(w=0\), since the uncertainty on each entry of Table~\ref{tab:hcw} is comparable to the variation across the table. That the total shift is small is in any case unsurprising: the cubic and \(O(3)\) fixed points are extremely close in the space of couplings, which is precisely the circumstance that makes the cubic theory difficult to isolate by other means.

\subsection{Operator dimensions}
Having fixed \(h_{\rm opt}(N)\) at each system size, we extract the scaling dimensions of the tracked operators from the DMRG spectrum at \(w=0.6\). The energy spectrum is normalised so that the vacuum energy vanishes and \(\Delta_\phi=0.5189\) in agreement with the Monte Carlo determination of \cite{Hasenbusch:2022zur}; all remaining scaled energies then furnish estimates of the corresponding CFT operator dimensions. The results for seven operators across system sizes \(N=12,\ldots,22\) are collected in Table~\ref{tab:dims}.

\begin{table}[ht]
\centering
\begin{tabular}{c|ccccccc}
    \(N\) & \(\Delta_\phi\) & \(\Delta_X\) & \(\Delta_Z\) & \(\Delta_S\) & \(\Delta_{A_\mu}\) & \(\Delta_{T_{\mu\nu}}\) & \(\Delta_{S'}\) \\
    \hline
    12 & 0.5189 & 1.2656 & 1.2467 & 1.5698 & 1.8781 & 2.9880 & 3.2622 \\
    14 & 0.5189 & 1.2631 & 1.2447 & 1.5831 & 1.9026 & 2.9969 & 3.2446 \\
    16 & 0.5189 & 1.2608 & 1.2428 & 1.5931 & 1.9202 & 3.0035 & 3.2325 \\
    18 & 0.5189 & 1.2588 & 1.2411 & 1.6004 & 1.9330 & 3.0077 & 3.2156 \\
    20 & 0.5189 & 1.2575 & 1.2399 & 1.6078 & 1.9439 & 3.0050 & 3.2048 \\
    22 & 0.5189 & 1.2553 & 1.2380 & 1.6122 & 1.9505 & 3.0049 & 3.1860 \\
\end{tabular}
\caption{Scaling dimensions of tracked operators at \(w=0.6\) as a function of system size \(N\). The value of \(\Delta_\phi\) is fixed by the normalisation of the spectrum. Values at \(N=12\) are obtained with ED; values at \(N=14,16,18\) are obtained with DMRG; values at \(N=20,22\) use DMRG with the extrapolated critical parameter \(h_{\rm opt}(N)\) from Table~\ref{tab:hopt}.}
\label{tab:dims}
\end{table}

The operators in Table~\ref{tab:dims} are identified by their quantum numbers in the fuzzy sphere spectrum. The fundamental scalar \(\phi\) (degeneracy 3) transforms in the three-dimensional fundamental representation of \(O_h\). The scalar singlet \(S\) (degeneracy 1) has \(l=0\) and trivial \(O_h\) representation; it is the leading relevant operator at the cubic fixed point and is the analogue of the thermal operator at the \(O(3)\) fixed point. The scalars \(X\) (degeneracy 2) and \(Z\) (degeneracy 3) arise from the decomposition of the five-dimensional \(l=2\) multiplet of \(O(3)\) under the cubic subgroup: under \(O_h\) this multiplet decomposes as \(E_g\oplus T_{2g}\), yielding a two-dimensional component (\(X\)) and a three-dimensional component (\(Z\)). The operator \(A_\mu\) is a spin-one primary (\(l=1\) on the sphere) that transforms as a vector under \(O(3)\), related to the \(O(3)\) Noether current at the isotropic fixed point. The stress-energy tensor \(T_{\mu\nu}\) is the unique spin-two (\(l=2\)) singlet primary, and \(S'\) is the next scalar in the singlet channel.

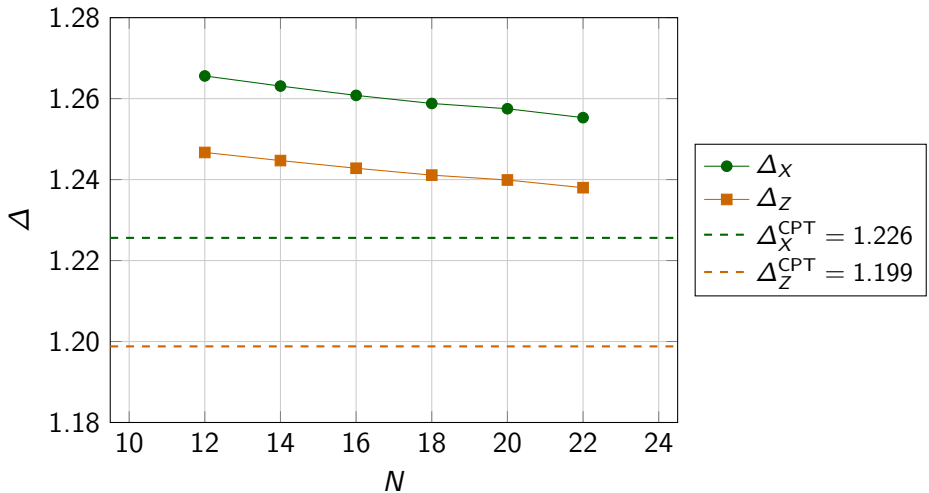
\begin{figure}[H]
    \centering
    \begin{tikzpicture}
        \begin{axis}[
            font=\sffamily\sansmath,
            width=0.55\textwidth,
            height=0.42\textwidth,
            xlabel={\(N\)},
            ylabel={\(\Delta\)},
            xmin=9.5, xmax=24.5,
            ymin=1.18, ymax=1.28,
            xtick={10,12,14,16,18,20,22,24},
            ytick distance=0.02,
            y tick label style={/pgf/number format/.cd, fixed, fixed zerofill, precision=2},
            grid=both,
            grid style={line width=0.2pt, draw=gray!30},
            major grid style={line width=0.4pt, draw=gray!40},
            tick align=inside,
            legend style={at={(1.03,0.5)},anchor=west,font=\sffamily\sansmath\small},
            legend cell align=left,
            ]
            \addplot[thin, green!40!black, mark=*, mark size=2pt]
              coordinates {(12,1.2656) (14,1.2631) (16,1.2608) (18,1.2588) (20,1.2575) (22,1.2553)};
            \addlegendentry{\(\Delta_X\)}
            \addplot[thin, orange!80!black, mark=square*, mark size=2pt]
              coordinates {(12,1.2467) (14,1.2447) (16,1.2428) (18,1.2411) (20,1.2399) (22,1.2380)};
            \addlegendentry{\(\Delta_Z\)}
            \addplot[dashed, thick, green!40!black, domain=9.5:24.5, samples=2, forget plot] {1.2256};
            \addplot[dashed, thick, green!40!black] coordinates {(0,0)(0,0)};
            \addlegendentry{\(\Delta_X^{\rm CPT}=1.226\)}
            \addplot[dashed, thick, orange!80!black, domain=9.5:24.5, samples=2, forget plot] {1.1988};
            \addplot[dashed, thick, orange!80!black] coordinates {(0,0)(0,0)};
            \addlegendentry{\(\Delta_Z^{\rm CPT}=1.199\)}
        \end{axis}
    \end{tikzpicture}
    \caption{Finite-size dependence of \(\Delta_X\) and \(\Delta_Z\) at \(w=0.6\). Both dimensions decrease monotonically towards the conformal perturbation theory predictions of \cite{Rong:2023owx} (dashed lines).}
    \label{fig:XZ}
\end{figure}

\paragraph{Scalars \boldmath{\(X\)} and \boldmath{\(Z\)}.}
Both \(\Delta_X\) and \(\Delta_Z\) decrease monotonically as \(N\) increases from 12 to 22, as illustrated in Fig.~\ref{fig:XZ}. At \(N=22\) we obtain \(\Delta_X=1.2553\) and \(\Delta_Z=1.2380\). These values lie above the conformal perturbation theory predictions \(\Delta_X^{\rm CPT}=1.2256(36)\) and \(\Delta_Z^{\rm CPT}=1.1988(24)\) of \cite{Rong:2023owx}, but the monotonically decreasing trends are consistent with eventual convergence towards those values. The non-degeneracy of \(X\) and \(Z\) is a direct consequence of the broken \(O(3)\) symmetry: at the \(O(3)\) fixed point these operators are degenerate members of the same five-dimensional multiplet, while the cubic deformation lifts this degeneracy. Across the range of system sizes studied the splitting is \(\Delta_X-\Delta_Z\approx 0.017\text{-}0.019\), somewhat below the conformal perturbation theory prediction \(\Delta_X^{\rm CPT}-\Delta_Z^{\rm CPT}\approx 0.027\), indicating that the spectrum has not yet fully converged.

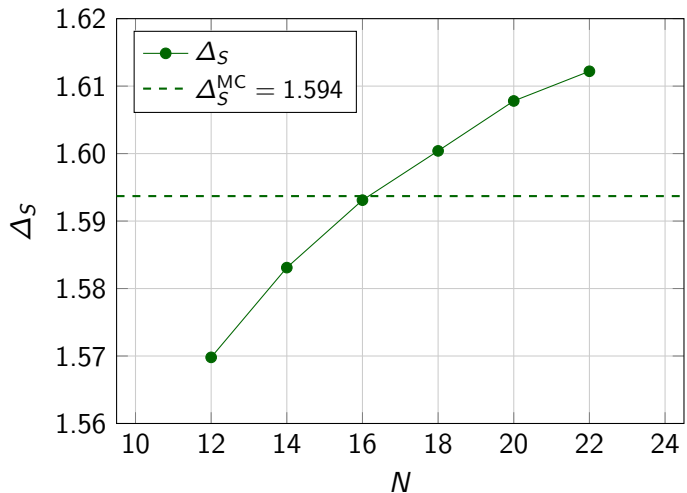
\begin{figure}[ht]
    \centering
    \begin{tikzpicture}
        \begin{axis}[
            font=\sffamily\sansmath,
            width=0.55\textwidth,
            height=0.42\textwidth,
            xlabel={\(N\)},
            ylabel={\(\Delta_S\)},
            xmin=9.5, xmax=24.5,
            ymin=1.56, ymax=1.62,
            xtick={10,12,14,16,18,20,22,24},
            ytick distance=0.01,
            y tick label style={/pgf/number format/.cd, fixed, fixed zerofill, precision=2},
            grid=both,
            grid style={line width=0.2pt, draw=gray!30},
            major grid style={line width=0.4pt, draw=gray!40},
            tick align=inside,
            legend style={at={(0.03,0.97)},anchor=north west,font=\sffamily\sansmath\small},
            legend cell align=left,
            ]
            \addplot[thin, green!40!black, mark=*, mark size=2pt]
              coordinates {(12,1.5698) (14,1.5831) (16,1.5931) (18,1.6004) (20,1.6078) (22,1.6122)};
            \addlegendentry{\(\Delta_S\)}
            \addplot[dashed, thick, green!40!black, domain=9.5:24.5, samples=2, forget plot] {1.5937};
            \addplot[dashed, thick, green!40!black] coordinates {(0,0)(0,0)};
            \addlegendentry{\(\Delta_S^{\rm MC}=1.594\)}
        \end{axis}
    \end{tikzpicture}
    \caption{Finite-size dependence of \(\Delta_S\) at \(w=0.6\). The dashed line shows the Monte Carlo determination \(\Delta_S=1.5937(6)\) of~\cite{Hasenbusch:2022zur}. The data increase monotonically with \(N\), crossing the reference near \(N=16\) and overshooting at larger system sizes.}
    \label{fig:S}
\end{figure}

\paragraph{Leading scalar singlet \boldmath{\(S\)}.}
The operator \(S\) is the first scalar singlet of \(O_h\) with \(l=0\) and degeneracy 1. It controls the approach to the cubic critical point and is the analogue of the thermal operator at the \(O(3)\) fixed point. The finite-size dependence of \(\Delta_S\) is shown in Fig.~\ref{fig:S}. 
The values increase monotonically from \(\Delta_S=1.5698\) at \(N=12\) to \(\Delta_S=1.6122\) at \(N=22\). The data pass through the Monte Carlo value \(\Delta_S^{\rm MC}=1.5937(6)\)~\cite{Hasenbusch:2022zur} between \(N=16\) (\(\Delta_S=1.5931\)) and \(N=18\) (\(\Delta_S=1.6004\)), and then overshoot it, ending approximately \(1.9\%\) above the reference at \(N=22\). This monotonically increasing behaviour, with values exceeding the MC benchmark for \(N\gtrsim 16\), is qualitatively different from the monotonically decreasing trends observed for \(X\) and \(Z\), and a reliable extrapolation to the thermodynamic limit is not attempted here.

\begin{figure}[ht]
    \centering
    \begin{tikzpicture}
        \begin{axis}[
            font=\sffamily\sansmath,
            width=0.55\textwidth,
            height=0.42\textwidth,
            xlabel={\(N\)},
            ylabel={\(\Delta_{T_{\mu\nu}}\)},
            xmin=9.5, xmax=24.5,
            ymin=2.98, ymax=3.015,
            xtick={10,12,14,16,18,20,22,24},
            ytick distance=0.005,
            y tick label style={/pgf/number format/.cd, fixed, fixed zerofill, precision=3},
            grid=both,
            grid style={line width=0.2pt, draw=gray!30},
            major grid style={line width=0.4pt, draw=gray!40},
            tick align=inside,
            legend style={at={(0.97,0.25)},anchor=north east,font=\sffamily\sansmath\small},
            legend cell align=left,
            ]
            \addplot[thin, green!40!black, mark=*, mark size=2pt]
              coordinates {(12,2.9880) (14,2.9969) (16,3.0035) (18,3.0077) (20,3.0050) (22,3.0049)};
            \addlegendentry{\(\Delta_{T_{\mu\nu}}\)}
            \addplot[dashed, thick, green!40!black, domain=9.5:24.5, samples=2, forget plot] {3.0};
            \addplot[dashed, thick, green!40!black] coordinates {(0,0)(0,0)};
            \addlegendentry{\(\Delta_{T_{\mu\nu}}=3\)}
        \end{axis}
    \end{tikzpicture}
    \caption{Finite-size dependence of \(\Delta_{T_{\mu\nu}}\) at \(w=0.6\). The dashed line marks the exact protected value \(\Delta_{T_{\mu\nu}}=d=3\).}
    \label{fig:Tmunu}
\end{figure}
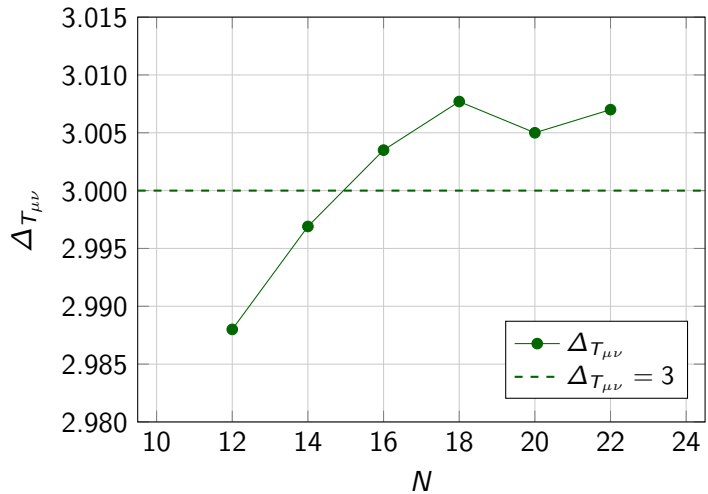

\paragraph{Stress-energy tensor \boldmath{\(T_{\mu\nu}\)}.}
In any local \(d\)-dimensional CFT the stress-energy tensor carries the protected dimension \(\Delta_{T_{\mu\nu}}=d\). For \(d=3\) this requires \(\Delta_{T_{\mu\nu}}=3\), and the fuzzy sphere data are consistent with this exact result, as shown in Fig.~\ref{fig:Tmunu}: the extracted values lie between \(2.988\) and \(3.008\), all within one percent of \(3\). This serves as a useful internal consistency check on the normalisation of the spectrum. We expect the mild non-monotonic behaviour at \(N\geq 20\) to settle to \(\Delta_{T_{\mu\nu}}=3\) at larger system sizes.

\begin{figure}[H]
    \centering
    \begin{tikzpicture}
        \begin{axis}[
            font=\sffamily\sansmath,
            width=0.55\textwidth,
            height=0.42\textwidth,
            xlabel={\(N\)},
            ylabel={\(\Delta_{A_\mu}\)},
            xmin=9.5, xmax=24.5,
            ymin=1.86, ymax=2.02,
            xtick={10,12,14,16,18,20,22,24},
            ytick distance=0.02,
            y tick label style={/pgf/number format/.cd, fixed, fixed zerofill, precision=2},
            grid=both,
            grid style={line width=0.2pt, draw=gray!30},
            major grid style={line width=0.4pt, draw=gray!40},
            tick align=inside,
            legend style={at={(0.03,0.73)},anchor=north west,font=\sffamily\sansmath\small},
            legend cell align=left,
            ]
            \addplot[thin, green!40!black, mark=*, mark size=2pt]
              coordinates {(12,1.8781) (14,1.9026) (16,1.9202) (18,1.9330) (20,1.9439) (22,1.9505)};
            \addlegendentry{\(\Delta_{A_\mu}\)}
            \addplot[dashed, thick, green!40!black, domain=9.5:24.5, samples=2, forget plot] {2.0};
            \addplot[dashed, thick, green!40!black] coordinates {(0,0)(0,0)};
            \addlegendentry{\(\Delta_{J_\mu}^{O(3)}=2\)}
        \end{axis}
    \end{tikzpicture}
    \caption{Finite-size dependence of \(\Delta_{A_\mu}\) at \(w=0.6\). The dashed line marks the protected dimension \(\Delta_{J_\mu}=d-1=2\) of the \(O(3)\) Noether current.}
    \label{fig:Amu}
\end{figure}
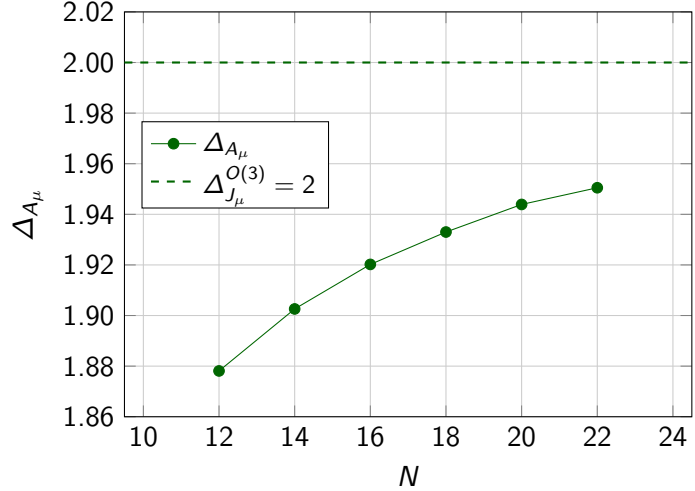

\paragraph{Vector operator \boldmath{\(A_\mu\)}.}
The spin-one operator \(A_\mu\) shows an increasing trend with \(N\), rising from \(\Delta_{A_\mu}=1.8781\) at \(N=12\) to \(1.9505\) at \(N=22\), as shown in Fig.~\ref{fig:Amu}. In the \(O(3)\) CFT the Noether current carries the protected dimension \(\Delta_{J_\mu}=d-1=2\). At the cubic fixed point the continuous \(O(3)\) symmetry is broken to the discrete group \(O_h\), rendering the would-be Noether currents non-conserved generic spin-one operators. The data are consistent with an extrapolated dimension close to but below \(2\), reflecting the proximity of the cubic and \(O(3)\) fixed points, though a precise determination requires larger system sizes.

\begin{figure}[ht]
    \centering
    \begin{tikzpicture}
        \begin{axis}[
            font=\sffamily\sansmath,
            width=0.55\textwidth,
            height=0.42\textwidth,
            xlabel={\(N\)},
            ylabel={\(\Delta_{S'}\)},
            xmin=9.5, xmax=24.5,
            ymin=2.97, ymax=3.30,
            xtick={10,12,14,16,18,20,22,24},
            ytick distance=0.05,
            y tick label style={/pgf/number format/.cd, fixed, fixed zerofill, precision=2},
            grid=both,
            grid style={line width=0.2pt, draw=gray!30},
            major grid style={line width=0.4pt, draw=gray!40},
            tick align=inside,
            legend style={at={(0.43,0.5)},anchor=north east,font=\sffamily\sansmath\small},
            legend cell align=left,
            ]
            \addplot[thin, green!40!black, mark=*, mark size=2pt]
              coordinates {(12,3.2622) (14,3.2446) (16,3.2325) (18,3.2156) (20,3.2048) (22,3.1860)};
            \addlegendentry{\(\Delta_{S'}\)}
            \addplot[dashed, thick, green!40!black, domain=9.5:24.5, samples=2, forget plot] {3.0133};
            \addplot[dashed, thick, green!40!black] coordinates {(0,0)(0,0)};
            \addlegendentry{\(\Delta_{S'}^{\rm MC}=3.013\)}
        \end{axis}
    \end{tikzpicture}
    \caption{Finite-size dependence of \(\Delta_{S'}\) at \(w=0.6\). The dashed line shows the Monte Carlo determination \(\Delta_{S'}=3.0133(8)\) of \cite{Hasenbusch:2022zur}. The data decrease monotonically with \(N\) but remain substantially above the reference value across the accessible system sizes.}
    \label{fig:epsprime}
\end{figure}
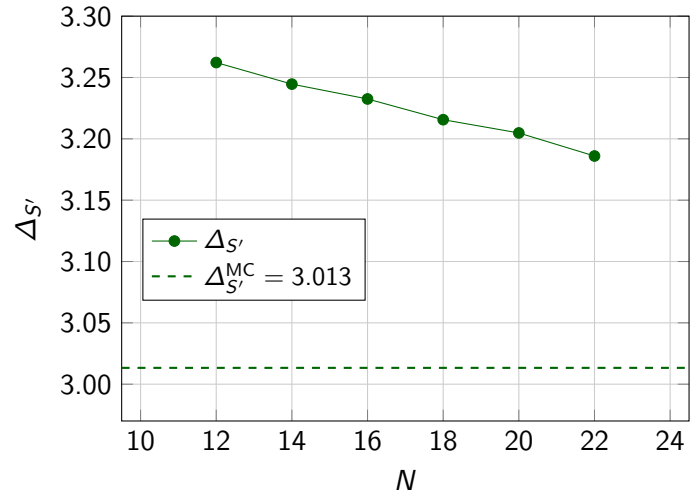

\paragraph{Second scalar singlet \boldmath{\(S'\)}.}
The dimension \(\Delta_{S'}\) decreases steadily from \(3.2622\) at \(N=12\) to \(3.1860\) at \(N=22\), as shown in Fig.~\ref{fig:epsprime}. These values lie well above the Monte Carlo result \(\Delta_{S'}=3.0133(8)\) of \cite{Hasenbusch:2022zur} and the conformal perturbation theory result \(\Delta_{S'}=3.0115(21)\) of \cite{Rong:2023owx}. Although the trend is in the correct direction, the rate of decrease suggests that considerably larger system sizes would be required to approach the reference values, and a reliable numerical extrapolation is not attempted here.

\subsection{Evidence for conformal symmetry and consistency of \texorpdfstring{\(h_c\)}{h\_c}}\label{sec:conformal}
The splitting of the \(l=2\) multiplet into \(X\) and \(Z\) demonstrates that the \(O(3)\) symmetry is broken to \(O_h\), but on its own it does not establish that the system sits at a critical point: an explicit breaking of \(O(3)\) down to \(O_h\) splits the quintet into a doublet and a triplet whether or not the theory is critical. The evidence that the Hamiltonian tuned to \(h_{\rm opt}(N)\) realises a conformal field theory is of a different and independent nature, and rests on the near-integer level spacings of the operators tracked here, to which we now turn.

In a CFT the states organise into multiplets consisting of a primary \(\mathcal{O}\) of dimension \(\Delta_{\mathcal{O}}\) together with its descendants, the first of which, \(\partial_\mu\mathcal{O}\), has dimension \(\Delta_{\mathcal{O}}+1\) and carries one additional unit of angular momentum on the sphere. Because the pair \((\phi,\partial_\mu\phi)\) is used to fix \(h_{\rm opt}\), the spacing of that particular tower is guaranteed by construction and carries no information. The towers built on the remaining operators, by contrast, are genuine predictions of conformal symmetry.

Level spacings \(\Delta_{\partial_\mu\mathcal{O}}-\Delta_{\mathcal{O}}\) are collected in Table~\ref{tab:towers}, over the same range of system sizes as Table~\ref{tab:dims}. All four approach the exact value of unity monotonically as \(N\) increases, the singlet tower from below and the \(X\) and \(Z\) towers from above; at \(N=22\) the deviations are \(1.4\%\) for \(S\) and close to \(3\%\) for \(X\) and \(Z\). The tower built on the vector \(A_\mu\) converges more slowly, improving from \(0.839\) at \(N=12\) to \(0.960\) at \(N=22\), consistently with the slower convergence of \(\Delta_{A_\mu}\) itself observed in Section~\ref{sec:results}.

\begin{table}[H]
\centering
\begin{tabular}{c|cccc}
    \(N\) & \(S\) & \(Z\) & \(X\) & \(A_\mu\) \\\hline
    12 & 0.9695 & 1.0354 & 1.0404 & 0.8387 \\
    14 & 0.9755 & 1.0352 & 1.0396 & 0.8740 \\
    16 & 0.9797 & 1.0343 & 1.0382 & 0.9024 \\
    18 & 0.9828 & 1.0328 & 1.0363 & 0.9250 \\
    20 & 0.9851 & 1.0317 & 1.0350 & 0.9449 \\
    22 & 0.9862 & 1.0299 & 1.0329 & 0.9601 \\
\end{tabular}
\caption{Conformal tower spacings \(\Delta_{\partial_\mu\mathcal{O}}-\Delta_{\mathcal{O}}\) between each primary \(\mathcal{O}\) and its first descendant, at \(w=0.6\), for the operators indicated. The exact value in a CFT is \(1\). The tower built on \(\phi\) is omitted, since the pair \((\phi,\partial_\mu\phi)\) is used to determine \(h_{\rm opt}\) and its spacing is therefore fixed by construction.}
\label{tab:towers}
\end{table}

These spacings also bear directly on the determination of \(h_c\). As shown in \eqref{eq:tunecrit}, fixing \(h_{\rm opt}\) amounts to requiring that a pair of gaps stand in the ratio of the corresponding CFT dimensions, a criterion that involves no OPE data and may be applied to any pair of operators whose dimensions are known independently. For a primary together with its own first descendant it takes a particularly simple form: because \(\Delta_{\partial_\mu\mathcal{O}}=\Delta_{\mathcal{O}}+1\) exactly, the condition \(\delta E_{\partial_\mu\mathcal{O}}/\delta E_{\mathcal{O}}=\Delta_{\partial_\mu\mathcal{O}}/\Delta_{\mathcal{O}}\) is equivalent to
\begin{equation}\label{eq:altcrit}
    \Delta_{\partial_\mu\mathcal{O}}-\Delta_{\mathcal{O}} = \frac{\Delta_{\mathcal{O}}}{\Delta_{\mathcal{O}}^{\rm ref}}\,,
\end{equation}
where the left-hand side is the measured level spacing of Table~\ref{tab:towers}, the \(\Delta\)'s on the right are the measured dimensions of Table~\ref{tab:dims}, and \(\Delta_{\mathcal{O}}^{\rm ref}\) is an independently known value. The entries of Table~\ref{tab:towers} therefore already measure how far each such alternative criterion is from being met at the value of \(h_{\rm opt}\) fixed by \((\phi,\partial_\mu\phi)\).

Taking \(\Delta_X^{\rm ref}=1.2256(36)\) and \(\Delta_Z^{\rm ref}=1.1988(24)\) from \cite{Rong:2023owx}, and \(\Delta_S^{\rm ref}=1.5937(6)\) from \cite{Hasenbusch:2022zur}, the criterion \eqref{eq:altcrit} is violated by \(0.9\%\) for \(X\) and by \(0.3\%\) for \(Z\) at \(N=22\), the latter being comparable to the quoted uncertainty of \(\Delta_Z^{\rm ref}\) itself. Both figures are stable across the whole range of system sizes studied. Tuning on \(X\) or on \(Z\) rather than on \(\phi\) would therefore leave \(h_c\) essentially unchanged. The singlet \(S\) is the least consistent of the three, with a violation of \(2.5\%\) at \(N=22\) that grows slowly with \(N\); this mirrors the overshoot of \(\Delta_S\) beyond the Monte Carlo value already noted in Section~\ref{sec:results}.

\section{Conclusion}\label{sec:conclusion}
We have demonstrated that the fuzzy sphere method can successfully isolate the cubic CFT in three dimensions. The essential ingredient is the addition of a cubic-invariant two-body interaction to the \(O(3)\) quantum rotor Hamiltonian of~\cite{Dey:2025zgn}. Expressed via the projector formulation~\eqref{eq:projcubic}, this deformation takes the form \(\sum_{a=x,y,z} P^a \otimes P^a\) within the triplet sector, and is incorporated into the full Hamiltonian~\eqref{eq:cubicHam} through the matrix \(\bar{\mathcal{M}}^{\mathcal{C}}\) with coupling \(w\). Because the cubic symmetry group \(O_h\) is hard-coded into the Hamiltonian by construction, the symmetry enhancement that has hampered conformal bootstrap approaches to the cubic theory is entirely absent.

The most direct evidence that the \(O(3)\) symmetry is broken to \(O_h\) is the non-degeneracy of the operators \(X\) and \(Z\), which arise from the decomposition of the five-dimensional rank-two traceless symmetric multiplet of \(O(3)\) into the \(E_g\) and \(T_{2g}\) representations of \(O_h\). At the \(O(3)\) fixed point these operators are exactly degenerate; the cubic deformation lifts this degeneracy, and across all system sizes \(N=12\)-\(22\) we observe a stable, monotonically convergent splitting \(\Delta_X - \Delta_Z \approx 0.017\)-\(0.019\), with both dimensions trending towards the conformal perturbation theory predictions of~\cite{Rong:2023owx}. That the resulting theory is conformal is a separate question, and is established independently in Section~\ref{sec:conformal} by the near-integer spacings between the operators tracked here and their first descendants, which approach the exact value of unity monotonically with \(N\) for every tower not used in the tuning. The leading scalar singlet \(S\) shows a monotonically increasing dimension, crossing the Monte Carlo reference \(\Delta_S^{\rm MC}=1.5937(6)\) near \(N=16\) and reaching \(\Delta_S=1.6122\) at \(N=22\); the overshoot and the qualitatively different convergence pattern compared to \(X\) and \(Z\) suggest sizeable finite-size corrections for this operator. Complementing these results, the stress-energy tensor dimension remains within one percent of the exact value \(\Delta_{T_{\mu\nu}}=3\) throughout, providing a robust internal consistency check. The spin-one operator \(A_\mu\), which is the conserved Noether current at the \(O(3)\) fixed point but becomes a non-conserved spin-one primary at the cubic fixed point, shows a dimension approaching a value below \(d-1=2\), consistent with the expected loss of current conservation.

One operator that remains challenging is the second scalar singlet \(S'\). Our values of \(\Delta_{S'}\) decrease from \(3.26\) at \(N=12\) to \(3.19\) at \(N=22\), substantially above the Monte Carlo and conformal perturbation theory benchmarks near \(3.013\). The operator in the \(O(3)\) model that \(S'\) stems from is the leading rank-four traceless symmetric one. In the fuzzy sphere study of~\cite{Dey:2025zgn}, this operator exhibits comparably large deviations from the bootstrap results in the literature, indicating that the slow convergence is not specific to the cubic deformation but is an intrinsic feature of the truncated quantum rotor setup.

Several improvements and extensions present themselves. On the numerical side, more systematic strategies for fine-tuning the Hamiltonian parameters---for instance, via conformal generators~\cite{Fardelli:2024qla, *Fan:2024vcz, *Fardelli:2026zas}---would sharpen the extrapolations and improve the accuracy of operator dimensions at larger system sizes. The study of operator product expansion coefficients in the cubic theory and comparison with those in the \(O(3)\) model would also be desirable. On the physical side, an interesting direction is the study of line and other defects in the cubic theory, extending results of \cite{Hu:2023ghk, *Zhou:2023fqu, *Zhou:2024dbt, *Dedushenko:2024nwi}. The \(\varepsilon\) expansion analysis of~\cite{Pannell:2023pwz} has already explored line defect RG flows and defect CFTs for theories with cubic symmetry. A non-perturbative treatment using the fuzzy sphere would provide complementary information about the defect fixed points and their operator content.

More broadly, the present work adds to the evidence that the fuzzy sphere method can access universality classes that are difficult to isolate by other means. The three-dimensional landscape of CFTs contains many interacting fixed points whose properties remain incompletely understood, including theories with discrete and continuous symmetry groups, as well as theories with multiple relevant couplings---see \cite{Pelissetto:2000ek} and \cite{Osborn:2017ucf, *Osborn:2020cnf, *Pannell:2023tzc} for expectations based on the $\varepsilon$ expansion. As the fuzzy sphere, conformal bootstrap, Monte Carlo, and perturbative expansion methods continue to mature individually and in concert, the prospect of a comprehensive non-perturbative classification and detailed study of \(d=3\) universality classes becomes increasingly tangible.

\ack{The author is supported by the Royal Society under grant URF\textbackslash{}R1\textbackslash211417 and by STFC under grant ST/X000753/1. The numerical computations in this work have used King's College London's CREATE~\cite{CREATE} computing cluster. Claude Code and Gemini were both used in this work for coding development and claim verification. They also aided in the writing of this paper.}

\bibliography{fuzzy_cubic}

\ifx\mcitethebibliography\mciteundefinedmacro
  \let\mcitethebibliography\thebibliography
  \expandafter\let\csname endmcitethebibliography\endcsname\endthebibliography
  \def\mcitedefaultmidpunct{,~}
  \def\mcitedefaultendpunct{.}
  \def\mcitedefaultseppunct{;}
  \def\EndOfBibitem{}
  \def\mciteBstWouldAddEndPuncttrue{}
  \def\mciteBstWouldAddEndPunctfalse{}
  \def\mciteSetBstMidEndSepPunct#1#2#3{#2}
\fi
\begin{mcitethebibliography}{10}
\ifx\href\asklfhas\newcommand{\href}[2]{#2}\fi
\ifx\arxivref\asklfhas\newcommand{\arxivref}[2]{\href{https://arxiv.org/abs/#1}{#2}}\fi
\ifx\doiref\asklfhas\newcommand{\doiref}[2]{\href{https://doi.org/#1}{#2}}\fi
\parskip 0pt
\normalsize

\bibitem{Rattazzi:2008pe}
R.~Rattazzi, V.~S. Rychkov, E.~Tonni \& A.~Vichi,
\textit{``{Bounding scalar operator dimensions in 4D CFT}''},
\doiref{10.1088/1126-6708/2008/12/031}{JHEP \textbf{0812}, 031
  (2008)\ignorespaces}\ignorespaces,
\texttt{\arxivref{0807.0004}{arXiv:0807.0004
  \![hep-th]}}\ignorespaces\mciteBstWouldAddEndPuncttrue
\mciteSetBstMidEndSepPunct{\mcitedefaultmidpunct\newline}
{\mcitedefaultendpunct}{\mcitedefaultseppunct}\relax
\EndOfBibitem\bibitem{Poland:2018epd}
D.~Poland, S.~Rychkov \& A.~Vichi,
\textit{``{The Conformal Bootstrap: Theory, Numerical Techniques, and
  Applications}''},
\doiref{10.1103/RevModPhys.91.015002}{Rev. Mod. Phys. \textbf{91}, 015002
  (2019)\ignorespaces}\ignorespaces,
\texttt{\arxivref{1805.04405}{arXiv:1805.04405
  \![hep-th]}}\ignorespaces\mciteBstWouldAddEndPuncttrue
\mciteSetBstMidEndSepPunct{\mcitedefaultmidpunct\newline}
{\mcitedefaultendpunct}{\mcitedefaultseppunct}\relax
\EndOfBibitem\bibitem{El-Showk:2012cjh}
S.~El-Showk, M.~F. Paulos, D.~Poland, S.~Rychkov, D.~Simmons-Duffin \&
  A.~Vichi,
\textit{``{Solving the 3D Ising Model with the Conformal Bootstrap}''},
\doiref{10.1103/PhysRevD.86.025022}{Phys. Rev. D \textbf{86}, 025022
  (2012)\ignorespaces}\ignorespaces,
\texttt{\arxivref{1203.6064}{arXiv:1203.6064
  \![hep-th]}}\ignorespaces\mciteBstWouldAddEndPuncttrue
\mciteSetBstMidEndSepPunct{\mcitedefaultmidpunct\newline}
{\mcitedefaultendpunct}{\mcitedefaultseppunct}\relax
\EndOfBibitem\bibitem{Kos:2014bka}
F.~Kos, D.~Poland \& D.~Simmons-Duffin,
\textit{``{Bootstrapping Mixed Correlators in the 3D Ising Model}''},
\doiref{10.1007/JHEP11(2014)109}{JHEP \textbf{1411}, 109
  (2014)\ignorespaces}\ignorespaces,
\texttt{\arxivref{1406.4858}{arXiv:1406.4858
  \![hep-th]}}\ignorespaces\mciteBstWouldAddEndPuncttrue
\mciteSetBstMidEndSepPunct{\mcitedefaultmidpunct\newline}
{\mcitedefaultendpunct}{\mcitedefaultseppunct}\relax
\EndOfBibitem\bibitem{Kos:2016ysd}
F.~Kos, D.~Poland, D.~Simmons-Duffin \& A.~Vichi,
\textit{``{Precision Islands in the Ising and $O(N)$ Models}''},
\doiref{10.1007/JHEP08(2016)036}{JHEP \textbf{1608}, 036
  (2016)\ignorespaces}\ignorespaces,
\texttt{\arxivref{1603.04436}{arXiv:1603.04436
  \![hep-th]}}\ignorespaces\mciteBstWouldAddEndPuncttrue
\mciteSetBstMidEndSepPunct{\mcitedefaultmidpunct\newline}
{\mcitedefaultendpunct}{\mcitedefaultseppunct}\relax
\EndOfBibitem\bibitem{Chang:2024whx}
C.-H. Chang, V.~Dommes, R.~S. Erramilli, A.~Homrich, P.~Kravchuk, A.~Liu, M.~S.
  Mitchell, D.~Poland \& D.~Simmons-Duffin,
\textit{``{Bootstrapping the 3d Ising stress tensor}''},
\doiref{10.1007/JHEP03(2025)136}{JHEP \textbf{2503}, 136
  (2025)\ignorespaces}\ignorespaces,
\texttt{\arxivref{2411.15300}{arXiv:2411.15300
  \![hep-th]}}\ignorespaces\mciteBstWouldAddEndPuncttrue
\mciteSetBstMidEndSepPunct{\mcitedefaultmidpunct\newline}
{\mcitedefaultendpunct}{\mcitedefaultseppunct}\relax
\EndOfBibitem\bibitem{Zhu:2022gjc}
W.~Zhu, C.~Han, E.~Huffman, J.~S. Hofmann \& Y.-C. He,
\textit{``{Uncovering Conformal Symmetry in the 3D Ising Transition:
  State-Operator Correspondence from a Quantum Fuzzy Sphere Regularization}''},
\doiref{10.1103/PhysRevX.13.021009}{Phys. Rev. X \textbf{13}, 021009
  (2023)\ignorespaces}\ignorespaces,
\texttt{\arxivref{2210.13482}{arXiv:2210.13482
  \![cond-mat.stat-mech]}}\ignorespaces\mciteBstWouldAddEndPuncttrue
\mciteSetBstMidEndSepPunct{\mcitedefaultmidpunct\newline}
{\mcitedefaultendpunct}{\mcitedefaultseppunct}\relax
\EndOfBibitem\bibitem{Zhou:2023qfi}
Z.~Zhou, L.~Hu, W.~Zhu \& Y.-C. He,
\textit{``{SO(5) Deconfined Phase Transition under the Fuzzy-Sphere Microscope:
  Approximate Conformal Symmetry, Pseudo-Criticality, and Operator
  Spectrum}''},
\doiref{10.1103/PhysRevX.14.021044}{Phys. Rev. X \textbf{14}, 021044
  (2024)\ignorespaces}\ignorespaces,
\texttt{\arxivref{2306.16435}{arXiv:2306.16435
  \![cond-mat.str-el]}}\ignorespaces\mciteBstWouldAddEndPuncttrue
\mciteSetBstMidEndSepPunct{\mcitedefaultmidpunct\newline}
{\mcitedefaultendpunct}{\mcitedefaultseppunct}\relax
\EndOfBibitem\bibitem{Zhou:2024zud}
Z.~Zhou \& Y.-C. He,
\textit{``{3D Conformal Field Theories with Sp(N) Global Symmetry on a Fuzzy
  Sphere}''},
\doiref{10.1103/xstj-xvcy}{Phys. Rev. Lett. \textbf{135}, 026504
  (2025)\ignorespaces}\ignorespaces,
\texttt{\arxivref{2410.00087}{arXiv:2410.00087
  \![hep-th]}}\ignorespaces\mciteBstWouldAddEndPuncttrue
\mciteSetBstMidEndSepPunct{\mcitedefaultmidpunct\newline}
{\mcitedefaultendpunct}{\mcitedefaultseppunct}\relax
\EndOfBibitem\bibitem{Fan:2025bhc}
R.~Fan, J.~Dong \& A.~Vishwanath,
\textit{``{Simulating the non-unitary Yang-Lee conformal field theory on the
  fuzzy sphere}''},
\texttt{\arxivref{2505.06342}{arXiv:2505.06342
  \![cond-mat.str-el]}}\ignorespaces\mciteBstWouldAddEndPuncttrue
\mciteSetBstMidEndSepPunct{\mcitedefaultmidpunct\newline}
{\mcitedefaultendpunct}{\mcitedefaultseppunct}\relax
\EndOfBibitem\bibitem{ArguelloCruz:2025zuq}
E.~Arguello~Cruz, I.~R. Klebanov, G.~Tarnopolsky \& Y.~Xin,
\textit{``{Yang-Lee Quantum Criticality in Various Dimensions}''},
\doiref{10.1103/w4qg-2xwn}{Phys. Rev. X \textbf{16}, 011022
  (2026)\ignorespaces}\ignorespaces,
\texttt{\arxivref{2505.06369}{arXiv:2505.06369
  \![hep-th]}}\ignorespaces\mciteBstWouldAddEndPuncttrue
\mciteSetBstMidEndSepPunct{\mcitedefaultmidpunct\newline}
{\mcitedefaultendpunct}{\mcitedefaultseppunct}\relax
\EndOfBibitem\bibitem{EliasMiro:2025msj}
J.~Elias~Mir{\'o} \& O.~Delouche,
\textit{``{Flowing from the Ising model on the fuzzy sphere to the 3D Lee-Yang
  CFT}''},
\doiref{10.1007/JHEP10(2025)037}{JHEP \textbf{2510}, 037
  (2025)\ignorespaces}\ignorespaces,
\texttt{\arxivref{2505.07655}{arXiv:2505.07655
  \![hep-th]}}\ignorespaces\mciteBstWouldAddEndPuncttrue
\mciteSetBstMidEndSepPunct{\mcitedefaultmidpunct\newline}
{\mcitedefaultendpunct}{\mcitedefaultseppunct}\relax
\EndOfBibitem\bibitem{He:2025ong}
Y.-C. He,
\textit{``{Free real scalar CFT on fuzzy sphere: spectrum, algebra and
  wavefunction ansatz}''},
\texttt{\arxivref{2506.14904}{arXiv:2506.14904
  \![hep-th]}}\ignorespaces\mciteBstWouldAddEndPuncttrue
\mciteSetBstMidEndSepPunct{\mcitedefaultmidpunct\newline}
{\mcitedefaultendpunct}{\mcitedefaultseppunct}\relax
\EndOfBibitem\bibitem{Taylor:2025odf}
J.~Taylor, C.~Voinea, Z.~Papi{\'c} \& R.~Fan,
\textit{``{Conformal Scalar Field Theory from Ising Tricriticality on the Fuzzy
  Sphere}''},
\doiref{10.1103/cj3l-cf58}{Phys. Rev. Lett. \textbf{136}, 056503
  (2026)\ignorespaces}\ignorespaces,
\texttt{\arxivref{2506.22539}{arXiv:2506.22539
  \![cond-mat.str-el]}}\ignorespaces\mciteBstWouldAddEndPuncttrue
\mciteSetBstMidEndSepPunct{\mcitedefaultmidpunct\newline}
{\mcitedefaultendpunct}{\mcitedefaultseppunct}\relax
\EndOfBibitem\bibitem{Zhou:2025rmv}
Z.~Zhou, C.~Wang \& Y.-C. He,
\textit{``{Chern-Simons-matter conformal field theory on fuzzy sphere:
  Confinement transition of Kalmeyer-Laughlin chiral spin liquid}''},
\texttt{\arxivref{2507.19580}{arXiv:2507.19580
  \![cond-mat.str-el]}}\ignorespaces\mciteBstWouldAddEndPuncttrue
\mciteSetBstMidEndSepPunct{\mcitedefaultmidpunct\newline}
{\mcitedefaultendpunct}{\mcitedefaultseppunct}\relax
\EndOfBibitem\bibitem{Voinea:2025iun}
C.~Voinea, W.~Zhu, N.~Regnault \& Z.~Papi{\'c},
\textit{``{Critical Majorana Fermion at a Topological Quantum Hall Bilayer
  Transition}''},
\doiref{10.1103/mztz-fyk3}{Phys. Rev. Lett. \textbf{136}, 076601
  (2026)\ignorespaces}\ignorespaces,
\texttt{\arxivref{2509.08036}{arXiv:2509.08036
  \![cond-mat.str-el]}}\ignorespaces\mciteBstWouldAddEndPuncttrue
\mciteSetBstMidEndSepPunct{\mcitedefaultmidpunct\newline}
{\mcitedefaultendpunct}{\mcitedefaultseppunct}\relax
\EndOfBibitem\bibitem{Tang:2025wtj}
Y.~Tang, C.~Voinea, L.~Hu, Z.~Papi{\'c} \& W.~Zhu,
\textit{``{Emergence of 3D Superconformal Ising Criticality on the Fuzzy
  Sphere}''},
\texttt{\arxivref{2512.25054}{arXiv:2512.25054
  \![cond-mat.str-el]}}\ignorespaces\mciteBstWouldAddEndPuncttrue
\mciteSetBstMidEndSepPunct{\mcitedefaultmidpunct\newline}
{\mcitedefaultendpunct}{\mcitedefaultseppunct}\relax
\EndOfBibitem\bibitem{Huffman:2026qqq}
E.~Huffman, Z.~Zhou, Y.-C. He \& J.~S. Hofmann,
\textit{``{Generalizing Deconfined Criticality to 3D $N$-Flavor
  $\mathrm{SU}(2)$ Quantum Chromodynamics on the Fuzzy Sphere}''},
\texttt{\arxivref{2602.11255}{arXiv:2602.11255
  \![hep-th]}}\ignorespaces\mciteBstWouldAddEndPuncttrue
\mciteSetBstMidEndSepPunct{\mcitedefaultmidpunct\newline}
{\mcitedefaultendpunct}{\mcitedefaultseppunct}\relax
\EndOfBibitem\bibitem{Dey:2026cso}
A.~Dey, L.~Herviou, C.~Mudry, S.~Rychkov \& A.~M. L{\"a}uchli,
\textit{``{Conformal Data for the $O(2)$ Wilson--Fisher CFT in
  $(2+1)$-Dimensional Spacetime from Exact Diagonalization and Matrix Product
  States on the Fuzzy Sphere}''},
\texttt{\arxivref{2604.18705}{arXiv:2604.18705
  \![cond-mat.str-el]}}\ignorespaces\mciteBstWouldAddEndPuncttrue
\mciteSetBstMidEndSepPunct{\mcitedefaultmidpunct\newline}
{\mcitedefaultendpunct}{\mcitedefaultseppunct}\relax
\EndOfBibitem\bibitem{Kos:2013tga}
F.~Kos, D.~Poland \& D.~Simmons-Duffin,
\textit{``{Bootstrapping the $O(N)$ vector models}''},
\doiref{10.1007/JHEP06(2014)091}{JHEP \textbf{1406}, 091
  (2014)\ignorespaces}\ignorespaces,
\texttt{\arxivref{1307.6856}{arXiv:1307.6856
  \![hep-th]}}\ignorespaces\mciteBstWouldAddEndPuncttrue
\mciteSetBstMidEndSepPunct{\mcitedefaultmidpunct\newline}
{\mcitedefaultendpunct}{\mcitedefaultseppunct}\relax
\EndOfBibitem\bibitem{Kos:2015mba}
F.~Kos, D.~Poland, D.~Simmons-Duffin \& A.~Vichi,
\textit{``{Bootstrapping the O(N) Archipelago}''},
\doiref{10.1007/JHEP11(2015)106}{JHEP \textbf{1511}, 106
  (2015)\ignorespaces}\ignorespaces,
\texttt{\arxivref{1504.07997}{arXiv:1504.07997
  \![hep-th]}}\ignorespaces\mciteBstWouldAddEndPuncttrue
\mciteSetBstMidEndSepPunct{\mcitedefaultmidpunct\newline}
{\mcitedefaultendpunct}{\mcitedefaultseppunct}\relax
\EndOfBibitem\bibitem{Chester:2020iyt}
S.~M. Chester, W.~Landry, J.~Liu, D.~Poland, D.~Simmons-Duffin, N.~Su \&
  A.~Vichi,
\textit{``{Bootstrapping Heisenberg magnets and their cubic instability}''},
\doiref{10.1103/PhysRevD.104.105013}{Phys. Rev. D \textbf{104}, 105013
  (2021)\ignorespaces}\ignorespaces,
\texttt{\arxivref{2011.14647}{arXiv:2011.14647
  \![hep-th]}}\ignorespaces\mciteBstWouldAddEndPuncttrue
\mciteSetBstMidEndSepPunct{\mcitedefaultmidpunct\newline}
{\mcitedefaultendpunct}{\mcitedefaultseppunct}\relax
\EndOfBibitem\bibitem{PhysRevB.110.115113}
C.~Han, L.~Hu \& W.~Zhu,
\textit{``Conformal operator content of the Wilson--Fisher transition on fuzzy
  sphere bilayers''},
\doiref{10.1103/PhysRevB.110.115113}{Phys. Rev. B \textbf{110}, 115113
  (2024)\ignorespaces}\ignorespaces,
\texttt{\arxivref{2312.04047}{arXiv:2312.04047
  \![cond-mat.str-el]}}\ignorespaces\mciteBstWouldAddEndPuncttrue
\mciteSetBstMidEndSepPunct{\mcitedefaultmidpunct\newline}
{\mcitedefaultendpunct}{\mcitedefaultseppunct}\relax
\EndOfBibitem\bibitem{Guo:2025odn}
W.~Guo, Z.~Zhou, T.-C. Wei \& Y.-C. He,
\textit{``{The $O(N)$ Free-Scalar and Wilson--Fisher Conformal Field Theories
  on the Fuzzy Sphere}''},
\texttt{\arxivref{2512.02234}{arXiv:2512.02234
  \![cond-mat.str-el]}}\ignorespaces\mciteBstWouldAddEndPuncttrue
\mciteSetBstMidEndSepPunct{\mcitedefaultmidpunct\newline}
{\mcitedefaultendpunct}{\mcitedefaultseppunct}\relax
\EndOfBibitem\bibitem{Dey:2025zgn}
A.~Dey, L.~Herviou, C.~Mudry \& A.~M. L{\"a}uchli,
\textit{``{Conformal Data for the O(3) Wilson--Fisher CFT from Fuzzy Sphere
  Realization of Quantum Rotor Model}''},
\texttt{\arxivref{2510.09755}{arXiv:2510.09755
  \![cond-mat.str-el]}}\ignorespaces\mciteBstWouldAddEndPuncttrue
\mciteSetBstMidEndSepPunct{\mcitedefaultmidpunct\newline}
{\mcitedefaultendpunct}{\mcitedefaultseppunct}\relax
\EndOfBibitem\bibitem{Pelissetto:2000ek}
A.~Pelissetto \& E.~Vicari,
\textit{``{Critical phenomena and renormalization group theory}''},
\doiref{10.1016/S0370-1573(02)00219-3}{Phys. Rept. \textbf{368}, 549
  (2002)\ignorespaces}\ignorespaces,
\texttt{\arxivref{cond-mat/0012164}{cond-mat/0012164}}\ignorespaces\mciteBstWouldAddEndPuncttrue
\mciteSetBstMidEndSepPunct{\mcitedefaultmidpunct\newline}
{\mcitedefaultendpunct}{\mcitedefaultseppunct}\relax
\EndOfBibitem\bibitem{Adzhemyan:2019gvv}
L.~T. Adzhemyan, E.~V. Ivanova, M.~V. Kompaniets, A.~Kudlis \& A.~I. Sokolov,
\textit{``{Six-loop $\varepsilon$ expansion study of three-dimensional
  $n$-vector model with cubic anisotropy}''},
\doiref{10.1016/j.nuclphysb.2019.02.001}{Nucl. Phys. B \textbf{940}, 332
  (2019)\ignorespaces}\ignorespaces,
\texttt{\arxivref{1901.02754}{arXiv:1901.02754
  \![cond-mat.stat-mech]}}\ignorespaces\mciteBstWouldAddEndPuncttrue
\mciteSetBstMidEndSepPunct{\mcitedefaultmidpunct\newline}
{\mcitedefaultendpunct}{\mcitedefaultseppunct}\relax
\EndOfBibitem\bibitem{Rong:2017cow}
J.~Rong \& N.~Su,
\textit{``{Scalar CFTs and Their Large N Limits}''},
\doiref{10.1007/JHEP09(2018)103}{JHEP \textbf{1809}, 103
  (2018)\ignorespaces}\ignorespaces,
\texttt{\arxivref{1712.00985}{arXiv:1712.00985
  \![hep-th]}}\ignorespaces\mciteBstWouldAddEndPuncttrue
\mciteSetBstMidEndSepPunct{\mcitedefaultmidpunct\newline}
{\mcitedefaultendpunct}{\mcitedefaultseppunct}\relax
\EndOfBibitem\bibitem{Stergiou:2018gjj}
A.~Stergiou,
\textit{``{Bootstrapping hypercubic and hypertetrahedral theories in three
  dimensions}''},
\doiref{10.1007/JHEP05(2018)035}{JHEP \textbf{1805}, 035
  (2018)\ignorespaces}\ignorespaces,
\texttt{\arxivref{1801.07127}{arXiv:1801.07127
  \![hep-th]}}\ignorespaces\mciteBstWouldAddEndPuncttrue
\mciteSetBstMidEndSepPunct{\mcitedefaultmidpunct\newline}
{\mcitedefaultendpunct}{\mcitedefaultseppunct}\relax
\EndOfBibitem\bibitem{Kousvos:2018rhl}
S.~R. Kousvos \& A.~Stergiou,
\textit{``{Bootstrapping Mixed Correlators in Three-Dimensional Cubic
  Theories}''},
\doiref{10.21468/SciPostPhys.6.3.035}{SciPost Phys. \textbf{6}, 035
  (2019)\ignorespaces}\ignorespaces,
\texttt{\arxivref{1810.10015}{arXiv:1810.10015
  \![hep-th]}}\ignorespaces\mciteBstWouldAddEndPuncttrue
\mciteSetBstMidEndSepPunct{\mcitedefaultmidpunct\newline}
{\mcitedefaultendpunct}{\mcitedefaultseppunct}\relax
\EndOfBibitem\bibitem{Kousvos:2019hgc}
S.~R. Kousvos \& A.~Stergiou,
\textit{``{Bootstrapping Mixed Correlators in Three-Dimensional Cubic Theories
  II}''},
\doiref{10.21468/SciPostPhys.8.6.085}{SciPost Phys. \textbf{8}, 085
  (2020)\ignorespaces}\ignorespaces,
\texttt{\arxivref{1911.00522}{arXiv:1911.00522
  \![hep-th]}}\ignorespaces\mciteBstWouldAddEndPuncttrue
\mciteSetBstMidEndSepPunct{\mcitedefaultmidpunct\newline}
{\mcitedefaultendpunct}{\mcitedefaultseppunct}\relax
\EndOfBibitem\bibitem{Kousvos:2025ext}
S.~R. Kousvos \& A.~Stergiou,
\textit{``{Redundancy channels in the conformal bootstrap}''},
\doiref{10.1007/JHEP01(2026)073}{JHEP \textbf{2601}, 073
  (2026)\ignorespaces}\ignorespaces,
\texttt{\arxivref{2507.05338}{arXiv:2507.05338
  \![hep-th]}}\ignorespaces\mciteBstWouldAddEndPuncttrue
\mciteSetBstMidEndSepPunct{\mcitedefaultmidpunct\newline}
{\mcitedefaultendpunct}{\mcitedefaultseppunct}\relax
\EndOfBibitem\bibitem{Sachdev:2011fcc}
S.~Sachdev,
\textit{``{Quantum Phase Transitions}''},
Cambridge University Press (2011)\ignorespaces\mciteBstWouldAddEndPuncttrue
\mciteSetBstMidEndSepPunct{\mcitedefaultmidpunct\newline}
{\mcitedefaultendpunct}{\mcitedefaultseppunct}\relax
\EndOfBibitem\bibitem{White:1992zz}
S.~R. White,
\textit{``{Density matrix formulation for quantum renormalization groups}''},
\doiref{10.1103/PhysRevLett.69.2863}{Phys. Rev. Lett. \textbf{69}, 2863
  (1992)\ignorespaces}\ignorespaces\mciteBstWouldAddEndPuncttrue
\mciteSetBstMidEndSepPunct{\mcitedefaultmidpunct\newline}
{\mcitedefaultendpunct}{\mcitedefaultseppunct}\relax
\EndOfBibitem\bibitem{White:1993zza}
S.~R. White,
\textit{``{Density-matrix algorithms for quantum renormalization groups}''},
\doiref{10.1103/PhysRevB.48.10345}{Phys. Rev. B \textbf{48}, 10345
  (1993)\ignorespaces}\ignorespaces\mciteBstWouldAddEndPuncttrue
\mciteSetBstMidEndSepPunct{\mcitedefaultmidpunct\newline}
{\mcitedefaultendpunct}{\mcitedefaultseppunct}\relax
\EndOfBibitem\bibitem{Fardelli:2024qla}
G.~Fardelli, A.~L. Fitzpatrick \& E.~Katz,
\textit{``{Constructing the Infrared Conformal Generators on the Fuzzy
  Sphere}''},
\doiref{10.21468/SciPostPhys.18.3.086}{Scipost Phys. \textbf{18}, 086
  (2025)\ignorespaces}\ignorespaces,
\texttt{\arxivref{2409.02998}{arXiv:2409.02998
  \![hep-th]}}\ignorespaces\mciteBstWouldAddEndPuncttrue
\mciteSetBstMidEndSepPunct{\mcitedefaultmidpunct\newline}
{\mcitedefaultendpunct}{\mcitedefaultseppunct}\relax
\EndOfBibitem\bibitem{Fan:2024vcz}
R.~Fan,
\textit{``{Note on explicit construction of conformal generators on the fuzzy
  sphere}''},
\texttt{\arxivref{2409.08257}{arXiv:2409.08257
  \![hep-th]}}\ignorespaces\mciteBstWouldAddEndPuncttrue
\mciteSetBstMidEndSepPunct{\mcitedefaultmidpunct\newline}
{\mcitedefaultendpunct}{\mcitedefaultseppunct}\relax
\EndOfBibitem\bibitem{Fardelli:2026zas}
G.~Fardelli, A.~L. Fitzpatrick \& E.~Katz,
\textit{``{Improving 3d Ising OPE Coefficients with Fuzzy Sphere Conformal
  Generators}''},
\texttt{\arxivref{2602.04958}{arXiv:2602.04958
  \![hep-th]}}\ignorespaces\mciteBstWouldAddEndPuncttrue
\mciteSetBstMidEndSepPunct{\mcitedefaultmidpunct\newline}
{\mcitedefaultendpunct}{\mcitedefaultseppunct}\relax
\EndOfBibitem\bibitem{Zhou:2025liv}
Z.~Zhou,
\textit{``{FuzzifiED : Julia Package for Numerics on the Fuzzy Sphere}''},
\texttt{\arxivref{2503.00100}{arXiv:2503.00100
  \![cond-mat.str-el]}}\ignorespaces\mciteBstWouldAddEndPuncttrue
\mciteSetBstMidEndSepPunct{\mcitedefaultmidpunct\newline}
{\mcitedefaultendpunct}{\mcitedefaultseppunct}\relax
\EndOfBibitem\bibitem{Haldane:1983xm}
F.~D.~M. Haldane,
\textit{``{Fractional quantization of the Hall effect: A Hierarchy of
  incompressible quantum fluid states}''},
\doiref{10.1103/PhysRevLett.51.605}{Phys. Rev. Lett. \textbf{51}, 605
  (1983)\ignorespaces}\ignorespaces\mciteBstWouldAddEndPuncttrue
\mciteSetBstMidEndSepPunct{\mcitedefaultmidpunct\newline}
{\mcitedefaultendpunct}{\mcitedefaultseppunct}\relax
\EndOfBibitem\bibitem{Antipin:2019vdg}
O.~Antipin \& J.~Bersini,
\textit{``{Spectrum of anomalous dimensions in hypercubic theories}''},
\doiref{10.1103/PhysRevD.100.065008}{Phys. Rev. D \textbf{100}, 065008
  (2019)\ignorespaces}\ignorespaces,
\texttt{\arxivref{1903.04950}{arXiv:1903.04950
  \![hep-th]}}\ignorespaces\mciteBstWouldAddEndPuncttrue
\mciteSetBstMidEndSepPunct{\mcitedefaultmidpunct\newline}
{\mcitedefaultendpunct}{\mcitedefaultseppunct}\relax
\EndOfBibitem\bibitem{Bednyakov:2023lfj}
A.~Bednyakov, J.~Henriksson \& S.~R. Kousvos,
\textit{``{Anomalous dimensions in hypercubic theories}''},
\doiref{10.1007/JHEP11(2023)051}{JHEP \textbf{2311}, 051
  (2023)\ignorespaces}\ignorespaces,
\texttt{\arxivref{2304.06755}{arXiv:2304.06755
  \![hep-th]}}\ignorespaces\mciteBstWouldAddEndPuncttrue
\mciteSetBstMidEndSepPunct{\mcitedefaultmidpunct\newline}
{\mcitedefaultendpunct}{\mcitedefaultseppunct}\relax
\EndOfBibitem\bibitem{Henriksson:2025hwi}
J.~Henriksson, F.~Herzog, S.~R. Kousvos \& J.~Roosmale~Nepveu,
\textit{``{Multiloop spectra in general scalar EFTs and CFTs}''},
\doiref{10.1016/j.physletb.2026.140235}{Phys. Lett. B \textbf{874}, 140235
  (2026)\ignorespaces}\ignorespaces,
\texttt{\arxivref{2507.12518}{arXiv:2507.12518
  \![hep-ph]}}\ignorespaces\mciteBstWouldAddEndPuncttrue
\mciteSetBstMidEndSepPunct{\mcitedefaultmidpunct\newline}
{\mcitedefaultendpunct}{\mcitedefaultseppunct}\relax
\EndOfBibitem\bibitem{Henriksson:2025vyi}
J.~Henriksson, S.~R. Kousvos \& J.~Roosmale~Nepveu,
\textit{``{EFT meets CFT: multiloop renormalization of higher-dimensional
  operators in general {\ensuremath{\phi}}$^{4}$ theories}''},
\doiref{10.1007/JHEP05(2026)005}{JHEP \textbf{2605}, 005
  (2026)\ignorespaces}\ignorespaces,
\texttt{\arxivref{2511.16740}{arXiv:2511.16740
  \![hep-th]}}\ignorespaces\mciteBstWouldAddEndPuncttrue
\mciteSetBstMidEndSepPunct{\mcitedefaultmidpunct\newline}
{\mcitedefaultendpunct}{\mcitedefaultseppunct}\relax
\EndOfBibitem\bibitem{Hasenbusch:2022zur}
M.~Hasenbusch,
\textit{``{Cubic fixed point in three dimensions: Monte Carlo simulations of
  the \(\phi^4\) model on the simple cubic lattice}''},
\doiref{10.1103/PhysRevB.107.024409}{Phys. Rev. B \textbf{107}, 024409
  (2023)\ignorespaces}\ignorespaces,
\texttt{\arxivref{2211.16170}{arXiv:2211.16170
  \![cond-mat.stat-mech]}}\ignorespaces\mciteBstWouldAddEndPuncttrue
\mciteSetBstMidEndSepPunct{\mcitedefaultmidpunct\newline}
{\mcitedefaultendpunct}{\mcitedefaultseppunct}\relax
\EndOfBibitem\bibitem{Carmona:1999rm}
J.~M. Carmona, A.~Pelissetto \& E.~Vicari,
\textit{``{The N component Ginzburg-Landau Hamiltonian with cubic anisotropy: A
  Six loop study}''},
\doiref{10.1103/PhysRevB.61.15136}{Phys. Rev. B \textbf{61}, 15136
  (2000)\ignorespaces}\ignorespaces,
\texttt{\arxivref{cond-mat/9912115}{cond-mat/9912115}}\ignorespaces\mciteBstWouldAddEndPuncttrue
\mciteSetBstMidEndSepPunct{\mcitedefaultmidpunct\newline}
{\mcitedefaultendpunct}{\mcitedefaultseppunct}\relax
\EndOfBibitem\bibitem{Rong:2023owx}
J.~Rong \& N.~Su,
\textit{``{From O(3) to Cubic CFT: Conformal Perturbation and the Large Charge
  Sector}''},
\texttt{\arxivref{2311.00933}{arXiv:2311.00933
  \![hep-th]}}\ignorespaces\mciteBstWouldAddEndPuncttrue
\mciteSetBstMidEndSepPunct{\mcitedefaultmidpunct\newline}
{\mcitedefaultendpunct}{\mcitedefaultseppunct}\relax
\EndOfBibitem\bibitem{fuzzy_cubic_code}
A.~Stergiou,
\textit{``{andstergiou/fuzzy-cubic: Zenodo archive for Quantum Rotors on the
  Fuzzy Sphere and the Cubic CFT}''},
\href{https://doi.org/10.5281/zenodo.20343464}{\nolinkurl{https://doi.org/10.5281/zenodo.20343464}}\mciteBstWouldAddEndPuncttrue
\mciteSetBstMidEndSepPunct{\mcitedefaultmidpunct\newline}
{\mcitedefaultendpunct}{\mcitedefaultseppunct}\relax
\EndOfBibitem\bibitem{Hu:2023ghk}
L.~Hu, Y.-C. He \& W.~Zhu,
\textit{``{Solving conformal defects in 3D conformal field theory using fuzzy
  sphere regularization}''},
\doiref{10.1038/s41467-024-47978-y}{Nature Commun. \textbf{15}, 3659
  (2024)\ignorespaces}\ignorespaces,
\texttt{\arxivref{2308.01903}{arXiv:2308.01903
  \![cond-mat.stat-mech]}}\ignorespaces,
[Erratum: \href{https://doi.org/10.1038/s41467-024-52959-2}{Nature Commun.
  \textbf{15}, 9013 (2024)}]\ignorespaces\mciteBstWouldAddEndPuncttrue
\mciteSetBstMidEndSepPunct{\mcitedefaultmidpunct\newline}
{\mcitedefaultendpunct}{\mcitedefaultseppunct}\relax
\EndOfBibitem\bibitem{Zhou:2023fqu}
Z.~Zhou, D.~Gaiotto, Y.-C. He \& Y.~Zou,
\textit{``{The $g$-function and defect changing operators from wavefunction
  overlap on a fuzzy sphere}''},
\doiref{10.21468/SciPostPhys.17.1.021}{SciPost Phys. \textbf{17}, 021
  (2024)\ignorespaces}\ignorespaces,
\texttt{\arxivref{2401.00039}{arXiv:2401.00039
  \![hep-th]}}\ignorespaces\mciteBstWouldAddEndPuncttrue
\mciteSetBstMidEndSepPunct{\mcitedefaultmidpunct\newline}
{\mcitedefaultendpunct}{\mcitedefaultseppunct}\relax
\EndOfBibitem\bibitem{Zhou:2024dbt}
Z.~Zhou \& Y.~Zou,
\textit{``{Studying the 3d Ising surface CFTs on the fuzzy sphere}''},
\doiref{10.21468/SciPostPhys.18.1.031}{SciPost Phys. \textbf{18}, 031
  (2025)\ignorespaces}\ignorespaces,
\texttt{\arxivref{2407.15914}{arXiv:2407.15914
  \![hep-th]}}\ignorespaces\mciteBstWouldAddEndPuncttrue
\mciteSetBstMidEndSepPunct{\mcitedefaultmidpunct\newline}
{\mcitedefaultendpunct}{\mcitedefaultseppunct}\relax
\EndOfBibitem\bibitem{Dedushenko:2024nwi}
M.~Dedushenko,
\textit{``{Ising BCFT from Fuzzy Hemisphere}''},
\texttt{\arxivref{2407.15948}{arXiv:2407.15948
  \![hep-th]}}\ignorespaces\mciteBstWouldAddEndPuncttrue
\mciteSetBstMidEndSepPunct{\mcitedefaultmidpunct\newline}
{\mcitedefaultendpunct}{\mcitedefaultseppunct}\relax
\EndOfBibitem\bibitem{Pannell:2023pwz}
W.~H. Pannell \& A.~Stergiou,
\textit{``{Line defect RG flows in the $\varepsilon$ expansion}''},
\doiref{10.1007/JHEP06(2023)186}{JHEP \textbf{2306}, 186
  (2023)\ignorespaces}\ignorespaces,
\texttt{\arxivref{2302.14069}{arXiv:2302.14069
  \![hep-th]}}\ignorespaces\mciteBstWouldAddEndPuncttrue
\mciteSetBstMidEndSepPunct{\mcitedefaultmidpunct\newline}
{\mcitedefaultendpunct}{\mcitedefaultseppunct}\relax
\EndOfBibitem\bibitem{Osborn:2017ucf}
H.~Osborn \& A.~Stergiou,
\textit{``{Seeking fixed points in multiple coupling scalar theories in the
  $\varepsilon$ expansion}''},
\doiref{10.1007/JHEP05(2018)051}{JHEP \textbf{1805}, 051
  (2018)\ignorespaces}\ignorespaces,
\texttt{\arxivref{1707.06165}{arXiv:1707.06165
  \![hep-th]}}\ignorespaces\mciteBstWouldAddEndPuncttrue
\mciteSetBstMidEndSepPunct{\mcitedefaultmidpunct\newline}
{\mcitedefaultendpunct}{\mcitedefaultseppunct}\relax
\EndOfBibitem\bibitem{Osborn:2020cnf}
H.~Osborn \& A.~Stergiou,
\textit{``{Heavy handed quest for fixed points in multiple coupling scalar
  theories in the $\varepsilon$ expansion}''},
\doiref{10.1007/JHEP04(2021)128}{JHEP \textbf{2104}, 128
  (2021)\ignorespaces}\ignorespaces,
\texttt{\arxivref{2010.15915}{arXiv:2010.15915
  \![hep-th]}}\ignorespaces\mciteBstWouldAddEndPuncttrue
\mciteSetBstMidEndSepPunct{\mcitedefaultmidpunct\newline}
{\mcitedefaultendpunct}{\mcitedefaultseppunct}\relax
\EndOfBibitem\bibitem{Pannell:2023tzc}
W.~H. Pannell \& A.~Stergiou,
\textit{``{Scalar-fermion fixed points in the {\ensuremath{\varepsilon}}
  expansion}''},
\doiref{10.1007/JHEP08(2023)128}{JHEP \textbf{2308}, 128
  (2023)\ignorespaces}\ignorespaces,
\texttt{\arxivref{2305.14417}{arXiv:2305.14417
  \![hep-th]}}\ignorespaces\mciteBstWouldAddEndPuncttrue
\mciteSetBstMidEndSepPunct{\mcitedefaultmidpunct\newline}
{\mcitedefaultendpunct}{\mcitedefaultseppunct}\relax
\EndOfBibitem\bibitem{CREATE}
{\relax King's College London},
\textit{``{King's Computational Research, Engineering and Technology
  Environment (CREATE)}''},
\href{https://doi.org/10.18742/rnvf-m076}{\nolinkurl{https://doi.org/10.18742/rnvf-m076}}\mciteBstWouldAddEndPuncttrue
\mciteSetBstMidEndSepPunct{\mcitedefaultmidpunct\newline}
{\mcitedefaultendpunct}{\mcitedefaultseppunct}\relax
\EndOfBibitem\end{mcitethebibliography}

\end{document}